\documentstyle[aps,preprint,epsf,tighten]{revtex}
\begin{document}
\draft
\title{ Universal  Correlations of  Coulomb Blockade
 Conductance Peaks and the Rotation Scaling in Quantum Dots}
\author {Y. Alhassid and H. Attias}
\address {Center for Theoretical Physics,  Sloane Physics Laboratory, 
Yale University, \\ New Haven, Connecticut 06520}
\date {Submitted October 1995}
\maketitle

\begin{abstract}
\narrowtext
 We show  that  the parametric correlations of the conductance peak amplitudes
 of a chaotic or weakly disordered quantum dot in the Coulomb blockade regime
 become universal upon an appropriate scaling of the parameter. 
 We compute the universal forms of this correlator  for both cases of 
conserved and 
broken time reversal symmetry. For a symmetric dot the correlator is
independent of the details in each lead such as the number of channels and
their correlation. We  derive a new scaling, which we call the rotation scaling,
 that can be computed directly from the dot's eigenfunction rotation rate
 or alternatively from the conductance peak heights,
 and therefore does not require knowledge of the spectrum of the dot. 
The relation of the rotation scaling to the level velocity scaling is discussed.
 The exact analytic form of the 
conductance peak correlator is derived at short distances. We also calculate the
 universal distributions of the average level  width velocity for various values
of the scaled parameter.
The universality is illustrated in an Anderson model of a disordered dot. 
\end{abstract}
\narrowtext

\section{Introduction}

A quantum dot \cite{DotsGeneral} is an isolated 2-D region 
 of a few $\mu$m or less in length
to which ${\cal N}$ electrons (typically several hundreds) are confined by an 
electrostatic potential.
The dots can be weakly coupled via tunnel barriers to external leads in order 
to study their transport properties. For sufficiently low temperatures the 
conductance of the dot exhibits equally spaced peaks with increasing gate 
voltage \cite{DotsExpt}. This phenomenon can be explained within a  model
\cite{CBtheory} where the leads are assumed to behave like 
electron reservoirs in thermal equilibrium at temperature $T$ and chemical 
potential $E_F$, and the dot electrons are treated as a free Fermi gas 
with the Coulomb interaction taken into account only through charging effects. 
In this framework each successive peak corresponds to a single-electron 
tunneling event. For an additional electron to tunnel from the leads into the 
dot and occupy the ground state of the resulting 
${\cal N}+1$-electron 
gas, it has to overcome the Coulomb repulsion energy of the dot electrons. 
 $E_F$ must  therefore increase from its value at the last tunneling
 event by  $\Delta E_F\approx e^2/C+\Delta$ where $C$ is the 
capacitance between the dot and the reservoir and 
$\Delta=E_{{\cal N}+1} - E_{\cal N}$ is the spacing between  
the single-particle energies $E_{\cal N}$. This 
condition provides a set of values $\{E_F\}$ at which tunneling is possible, 
resulting in peaks in the conductance $G(E_F)$ of the dot. The suppression of
tunneling between the peaks by Coulomb repulsion is called Coulomb blockade.

Since the charging energy $e^2/C\gg\Delta$, these peaks are equally
spaced as a function of $E_F$  (for $kT < e^2/C$) and can be 
observed by increasing the gate voltage which controls $E_F$.
 The resulting 
oscillations in the measured conductance are known as Coulomb blockade 
oscillations. In the regime of isolated resonances 
$\overline{\Gamma}_\lambda\ll\Delta$ where $\Gamma_\lambda$ is the decay
 rate 
from the single-particle level $E_\lambda$ into the leads, and for 
temperatures $kT  \ll \Delta$,  the conductance is 
dominated by the contribution of the single level closest to $E_F$. In
 the temperature regime 
 $kT\gg\overline{\Gamma}_\lambda$, the width of the
peaks is $\sim kT$  independent of the individual decay rates.

The peak amplitudes are found to fluctuate strongly \cite{DotsExpt}. These
fluctuations have been explained \cite{RMTforDots} by a statistical theory 
based on the assumption that the Hamiltonian of the dot can be described by 
random-matrix theory (RMT) \cite{RMTreviews,Billiard}. This assumption is 
valid either
in the diffusive regime (for weak disorder) or when the classical dynamics of
the electron inside the dot is chaotic \cite{Chaos}  due to irregularities in 
the confining 
potential.  The full probability distribution of the conductance was 
derived in this framework and found to change dramatically
when time reversal symmetry is broken due to the presence of a magnetic
field \cite{RMTforDots}.

 Recent experiments have been probing the dot's conductance as a function of
the external magnetic field or a changing shape. Of particular interest  are
the correlations of the conductance peaks at different values of the external
parameter  (magnetic field, shape, etc.).  For ballistic cavities (open dots),
 conductance  correlations in
a magnetic field were studied  extensively,  both theoretically for a large
number of channels  using the semiclassical
approximation \cite{DotsSemiCl}  and the supersymmetry method
 \cite{OpenDotCorr},
and  experimentally \cite{BallisDotsExpt}. Recently 
 these correlations were measured also as a function of the dot's shape 
\cite{ShapeExpt}.
  Such open dots are characterized by
 many overlapping resonances and exhibit properties that
are analogous to  Ericson's fluctuations in nuclear reactions \cite{EricFluct}. 
However, until recently, very little was known about parametric conductance
 correlations in the Coulomb blockade regime of isolated resonances,
 which is analogous to the neutron resonance regime in the compound nucleus 
\cite{Porter,RMTreviews}.
 
 In this paper we focus on the parametric correlation function of the
conductance peaks 
\begin{eqnarray}\label{gParmCorr}
   c_G(x-x^\prime)=\overline{{\tilde G}(x){\tilde G}(x^\prime)}
   \;,\;\;\;\;\;
   {\tilde G}(x)={G(x)-\overline{G}\over\sqrt{\overline{G^2}-\overline{G}^2}}
\end{eqnarray}
where $x$ is any controllable parameter and $G(x)$ is the conductance 
peak amplitude for a given value $x$ of the parameter. 
 In Ref. \cite{Cond} we have proposed and derived for the first time
 the universal  form of this correlator
 in the Coulomb blockade regime by using the framework 
of a Gaussian  process (GP)  \cite{GP,GPlong} that generalizes the Gaussian
 ensembles (GE) to situations where the chaotic or disordered Hamiltonian
 depends on a parameter. A  particular realization of a GP has been introduced 
in Ref. \cite{Wil} to study the statistics of avoided crossings in chaotic 
systems. Universal  parametric correlation of  spectra (e.g. level velocities)
 have been derived in Ref. \cite{SpectCorr,MatrixModels}. However, the 
calculation of the
 conductance correlations requires knowledge of statistical properties of
the eigenfunctions.
The universality of all eigenfunction correlators, which implies in particular 
the universality of (\ref{gParmCorr}), was demonstrated in \cite{GP}. 
  Following our predictions, the conductance correlator has recently been
measured for the situation of broken time-reversal symmetry  where $x$ is 
a magnetic field \cite{ExpCond} and was found to be in good agreement
with theory \cite{Cond}. The conductance distributions
for one-channel symmetric leads were also measured \cite{ExpDist,ExpCond}
 recently and are in accord with the theoretical distributions derived in 
Ref. \cite{RMTforDots}.

  For a dot with reflection symmetry we derive an even more remarkable
result: the conductance correlator is independent of the  details of the 
channels 
in each lead  such as their  number and degree of correlations, and it reduces
to the resonance width correlator for single-channel leads.
However, the conductance correlations are sensitive to the absence 
or presence of  a magnetic field. 
 To demonstrate the universality of our results we  test them in an
Anderson model of a  disordered  quantum dot. 

  Another problem that we address  is the difficulty in performing the usual
 level velocity parameter scaling in experimental situations
because of the inaccessibility of the spectrum, a Coulomb blockade effect. 
One can trace the variation of the ground state energy of ${\cal N}$ electrons 
in the dot as a function of the parameter by following the corresponding
 conductance peak.
However  it is not possible to measure the level spacing  $\Delta$ which
is needed for the level velocity scaling. Though $\Delta$ may be estimated
 theoretically, e.g. from free Fermi gas, its value may be modified by the
 electrons' interaction.  We find a new scaling procedure,
 with a scaling factor which
 depends on the eigenfunctions alone and can be extracted from the level
 width or the conductance peak\cite{Cond}. 
This  should make it possible to extract the scaling 
directly from the
 conductance peaks data.  The scaling factor is 
interpreted as the average rotation rate of the eigenvectors (with
 respect to $x$) or as the rms conductance velocity.

We remark that  the conductance correlator in a symmetric dot is
identical to the parametric correlator of the eigenfunction intensity
 at a fixed spatial point of a chaotic system, and therefore can also be
measured in microwave cavity experiments\cite{Microwave1,Microwave2}.

 The outline of the paper is as follows. In Section II we briefly review the 
theory of
conductance in quantum dots in their Coulomb blockade regime using 
$R$-matrix formalism \cite{R-Matrix,WigRmat,NucTheory},  
or alternatively -- resonance
 theory \cite{NucTheory}. Both formalisms
are particularly suitable for making the connection to RMT 
\cite{RMTforDots,RobnikDots,RobnikDots1}.  In Section III we 
derive and demonstrate the universal parametric correlations of the 
conductance peaks.  We consider symmetric 
dots and show that in this case
$c_G(x-x^\prime)$ is not only universal  upon an appropriate scaling of the 
external parameter, but also independent of the channel 
information. We compute its universal form using
 a simple GP for both cases of conserved
 and broken time reversal symmetry.  In section IV we introduce the new 
scaling procedure  based on the eigenfunction rotation rate.
 The scaling factor  diverges logarithmically for the orthogonal case 
 but we describe a way to  regularize it.
 Using perturbation theory, we  derive
 analytically the width correlator at  short distances and express
 the rotation scaling
in terms of the RMS of the width velocity.    We also 
derive a semiclassical expression  which is system-dependent 
for the rotation scaling and compute 
it for the case when the parameter is an external magnetic field. 
This gives us the magnitude of the correlation field.
In Section V
we compute the universal distributions of the average width velocity
 for various values of  the scaled $\Delta  x$.
In Section VI  we discuss the universal parametric correlations 
of the conductance
 peaks in the general case of asymmetric dots with symmetric or asymmetric
 leads, and express the rotation scaling in terms of the RMS of the conductance
velocity. Finally, our analytical results are verified in the completely 
solvable $N=2$  GP (Section VII). 

\section{Conductance in Quantum Dots}

\subsection{$R$-Matrix Formalism}

It is convenient to discuss the conductance of the dot in the framework of 
$R$-matrix theory, originally introduced  
for nuclear reactions that proceed through the compound nucleus formation
 and decay \cite{WigRmat,NucTheory,R-Matrix} and more recently applied to 
quantum dots \cite{RMTforDots,RobnikDots,RobnikDots1}.
 We briefly review this formalism in the context of quantum dots
 but for the general case where there are several open channels at each lead.
 Let us consider a  planar cavity-like dot 
in  the $x$-$y$ plane with left and right leads of 
width  $D_l,D_r$ attached to it at the lines of contact $x=x_l,x_r$; 
$0\leq y\leq D_{l,r}$ which we denote by ${\cal C}$. The dot Hamiltonian has 
energies $E_\lambda$ whose corresponding eigenfunctions
 $\Psi_\lambda(\vec{r})$ 
vanish at the walls and satisfy an homogeneous boundary
 condition 
\begin{eqnarray}\label{LeadsBC}
   \partial\Psi_\lambda/\partial n-h_{l,r}\Psi_\lambda=0
   \;\;\;\;\; 
   {\rm for}\;\vec{r}\in{\cal C}
\end{eqnarray}
where $\hat{n}$ is the
normal to ${\cal C}$ (i.e. $\hat{n}=\pm\hat{x}$) and $h_{l,r}$ are constants.  
 
We denote by $\phi^i_{c}$
a complete set of  transverse wavefunctions, where $c$ is a channel 
index and $i=l,r$. For an open channel 
$\phi^i_{c} (y)= \sqrt{2/D_i} \sin (\kappa^i_c y)$, 
where $\kappa^i_c=c \pi/D_i \; (c=1,2,\ldots,\Lambda_i)$
 is the quantized transverse momentum. 
 Inside each lead we can expand  the scattering solution $\Phi$ at
energy $E$  as
\begin{eqnarray}\label{PhiExpandChan}
   \Phi(\vec{r})=\sum\limits_{c=1}^{\Lambda_i}u_c^i(x)\phi_c^i(y) \;.
\end{eqnarray}
  The  $R$-matrix,  which relates the values of
 $\partial u^{ i^\prime}_{c^\prime}/\partial n - 
h_{i^\prime} u^{ i^\prime}_{c^\prime}$ to those of $u^i_{c}$ 
at  $x=x_i$
  can be expressed in terms of the resonances' eigenfunctions and energies 
through
   $R_{i^\prime c^\prime; ic}(E)=
     \sum\limits_\lambda
     y_{c\lambda}^i  y_{c^\prime\lambda}^{{i^\prime}\;\ast} /
   (E_\lambda-E)$
where 
\begin{eqnarray}\label{PartialWamp}
    y_{c\lambda}^i= \sqrt{ {\hbar^2 \over 2m} }
   \int\limits_{\cal C}dl\;
   \phi_c^{i\;\ast}\Psi_\lambda
\end{eqnarray}
is the reduced partial width amplitude for the decay from level 
$\lambda$ into channel $c$ in lead $i$. 

   If the resonances are isolated, i.e. their total width 
 (see below)  is much smaller than their average
 spacing $\Delta$, only one resonance whose energy $E_\lambda$ 
 is closest to $E$  contributes to the $R$-matrix,
and the cross section for scattering from 
channel $c$ in lead $i$ into channel $c^\prime$ in lead $i^\prime$ 
($i \neq i^\prime$) is 
given by the Breit-Wigner resonance formula    \cite{R-Matrix}
\begin{eqnarray}\label{BreitWigner}
   \mid S_{i^\prime c^\prime; ic}\mid^2=
   {\Gamma_{c\lambda}^i\Gamma_{c^\prime\lambda}^{i^\prime}\over
   (E-E_\lambda)^2+{1\over 4}(\Gamma_\lambda^l+\Gamma_\lambda^r)^2} \;.
\end{eqnarray}
 Here $\Gamma^i_{c\lambda}$ and $\gamma^i_{c\lambda}$ are respectively
 the partial width and partial width amplitude of the resonance level 
$\lambda$ to decay into
 channel $c$ in lead $i$
\begin{eqnarray}\label{PartialW}
\Gamma^i_{c\lambda} = \mid\gamma^i_{\lambda}\mid^2 \;;\;\;\;\;
\gamma^i_{c\lambda}= \sqrt{2 k_c^i P_c^i}  y^i_{c \lambda}  \;
\end{eqnarray}
 where $k_c^i$ is the longitudinal channel momentum
$\hbar^2k_c^{i 2}/2m + \hbar^2\kappa_c^{i 2}/2m = E$, 
and $P_c^i$ is the penetration factor to tunnel through
the barrier in channel $c$.  
 $\Gamma^i_\lambda = \sum_{c} \Gamma^i_{c\lambda}$ is the total decay
 width of level $\lambda$ into lead $i$.

  The conductance of the dot at energy $E$ is given by Landauer's formula
\cite{LandauerForm} 
   $g(E)=(e^2/ h)\sum_{cc^\prime}\mid S_{rc^\prime; lc}\mid^2$.
 At zero temperature (or for $kT\ll \Gamma_\lambda$) the conductance
 as a function of the Fermi energy is given by $g(E=E_F)$,
 while at finite temperature the conductance $g(E)$ due to one resonance
 level $\lambda$ has to be convoluted with
the derivative of the Fermi function 
$f(\epsilon)=\left(1+e^{\epsilon/kT}\right)^{-1}$ 
\begin{eqnarray}\label{gFiniteT}
   G(E_F)=-\int dE g(E)f^\prime(E-E_F) \;.
\end{eqnarray} 
In the temperature regime $\overline{\Gamma_\lambda}\ll kT\ll\Delta$ where
Coulomb-blockade oscillations have been observed \cite{DotsExpt}, the peak of 
$g(E)$ at $E=E_\lambda$ is much narrower than the peak of $-f^\prime(E-E_F)$ 
there and we can approximate (\ref{gFiniteT}) by
\begin{eqnarray}\label{GatCBregime}
   G(E_F)={e^2\over h}{\pi\over 2kT}
   {\Gamma_\lambda^l\Gamma_\lambda^r\over\Gamma_\lambda^l+
\Gamma_\lambda^r}
   \cosh^{-2}\left({E_F-E_\lambda\over 2kT}\right)
   \;.
\end{eqnarray}
The conductance has therefore a resonance shape as a function of $E_F$ 
centered
at the resonance energy $E_F=E_\lambda$ with peak amplitude of
\begin{eqnarray}\label{PeakG}
   G={e^2\over h}{\pi\over 2kT}
   {\Gamma_\lambda^l\Gamma_\lambda^r\over\Gamma_\lambda^l+
\Gamma_\lambda^r}
\end{eqnarray}     
and width of $kT$. This formula has been derived in \cite{CBtheory} using a
different method.

 We remark that a different  modelling of a quantum dot assumes point-like 
contacts where each lead is composed of several such point contacts.  
One arrives at the same  expression
 (\ref{PeakG}) for the conductance \cite{DotDistSUSY},
 except that each point contact ${\bf r}_c$ is considered as one channel 
and the corresponding reduced partial width is \cite{PEI}
$ \gamma_{c\lambda} =
 \sqrt{ {\alpha_c \cal A} \Delta/\pi} \Psi_\lambda({\bf r}_c)$, 
where  ${\cal A}$ is the area of the dot, $\Delta$ is the mean level 
spacing and $\alpha_c$ is a coupling parameter of the dot to the lead
that represents a tunneling probability through the barrier. 

\subsection{Resonance Theory}

Eqs.  (\ref{PartialW}) and (\ref{PartialWamp}), which  relate the partial 
 widths to the resonance eigenfunction
in the dot,  play a major role in the statistical theory for the conductance
peaks (\ref{PeakG}), and it is therefore instructive to derive them 
directly from resonance theory \cite{NucTheory}
 without using  the $R$-matrix formalism.

A decaying state is described 
by the condition that the wavefunction has only an outgoing component in
the external region (instead of the boundary condition  (\ref{LeadsBC})
 used in the $R$-matrix formalism). In our case a general solution $\Phi$
 at energy $E$ has inside the leads the form (\ref{PhiExpandChan}) with 
  $u_c(x) = A_c u_c^+(x) +B_cu_c^-(x)$
where  $u_c^\pm(x)$ are incoming and outgoing longitudinal waves
propagating through a potential barrier $V_c(x)$ in channel $c$.  The conditions
$B_c=0$ in all channels determine a discrete set of complex energies 
$E_\lambda^R= E_\lambda + \delta E_\lambda -i \Gamma_\lambda/2$, where
$\delta E_\lambda$ is a real shift in the resonance energy and 
$\Gamma_\lambda = \sum_c \Gamma_{c\lambda}$ is the total resonance
 width. An equivalent boundary condition for determining the complex resonance
energies is that the logarithmic derivative at the interface of the leads with
 the dot
is equal to the logarithmic derivative of the outgoing wave
\begin{eqnarray}\label{resonbc}
{\partial\over\partial n}  \ln u_c^i\mid_{x=x_i} = h^+_{ic} \equiv 
{\partial \over \partial n} \ln u^{i+}_c\mid_{x=x_i}\;.
\end{eqnarray}
 $h_{ic}^+$ are generally complex, e.g. $h_{ic}^+ =\pm ik^i_c$ in the absence 
of barrier
 (when $u_c^+$ is a plane wave propagating in a direction away from the dot). 
The modification of the boundary 
conditions from (\ref{LeadsBC}) (where $h_i$ are real) to (\ref{resonbc})
 allows the wavefunction to leak into 
the leads. The partial decay rate $\Gamma_{c\lambda}/\hbar$  into channel
 $c$ of lead $i$ is given
by  the current through the cross-section of the lead
\begin{eqnarray}
{\Gamma^i_{c\lambda} \over \hbar} = {\hbar \over 2im}\int\limits_{\cal C}dl
   \left(\Psi_\lambda^\ast{\partial\Psi_\lambda\over\partial n}
   -\Psi_\lambda{\partial\Psi_\lambda^\ast\over\partial n}\right)_c =
 {\hbar \over 2im} \left[ u_c^\ast {du_c \over dn} - u_c 
{du_c^\ast \over dn}\right]_{x=x_i}
\;.
\end{eqnarray}

 Using the boundary conditions (\ref{resonbc}) we find
\begin{eqnarray}
\Gamma^i_{c\lambda} = {\hbar^2 \over m}  {\rm Im} h_{ic}^+ 
 \mid u_c(x_i)\mid^2\;,
\end{eqnarray}
which is exactly Eqs. (\ref{PartialW}) and (\ref{PartialWamp}) provided the
 penetration factor $P_c$ is defined by 
\begin{eqnarray}\label{penet}
 k^i_cP^i_c \equiv {\rm Im} h_{ic}^+ \;.
\end{eqnarray}
 In the absence of a barrier  $P^i_c=1$, and we expect
$P^i_c \ll 1$ in the presence of a barrier. 

\subsection{Statistical Model for Dots with Multi-channel Leads}

Eq. (\ref{PeakG}) indicates that the irregular fluctuations of the conductance 
peaks arise from fluctuations of the
level widths $\Gamma_{c\lambda}^i$.  We assume that the channel penetration 
factors 
$P_c^i$ have negligible dependence on the parameter, and in any case they 
vary smoothly with $x$.  
These fluctuations must then come from the
eigenfunctions $\Psi_\lambda$ inside the dot, 
and have recently been accounted 
for by a statistical theory \cite{RMTforDots} based on the assumption that 
the Hamiltonian of the dot can be described by a $N\times N$ random matrix $H$
taken from the appropriate GE \cite{RMTreviews}. The dot's eigenfunctions 
$\Psi_\lambda$ are expanded in a complete set of solutions $\rho_\mu^\lambda$ 
 inside the dot with the given energy $E_\lambda$,  
 $ \Psi_\lambda(\vec{r})=\sum\limits_{\mu=1}^N \psi_\mu^\lambda 
   \rho_\mu^\lambda(\vec{r})$,
and the expansion coefficients $\psi_\mu^\lambda$ are assumed to constitute 
the components of the $\lambda$-th eigenvector of the random matrix $H$.
 The partial width amplitude to decay into channel $c$ (\ref{PartialWamp}) 
can then be expressed as a scalar product
\begin{eqnarray}\label{PartWampProj}
   \gamma_{c \lambda}^i=\sum\limits_{\mu=1}^N 
   \phi_{c\mu}^{i\ast} \psi^\lambda_\mu
   \equiv\langle\phi_c^i\mid\psi_\lambda\rangle \;,
  \end{eqnarray}
where $\phi_c^i$ is the channel vector defined by
 $ \phi_{c\mu}^i=\sqrt{\hbar^2k_c^iP_c^i /m}
   \int\limits_{\cal C}dl\;\phi_c^{i\;\ast}\rho_\mu$.
In the point contact model, a similar formula (\ref{PartWampProj}) applies
 but with $\phi_{c\mu} \equiv  \sqrt{\alpha_c{\cal A}\Delta/\pi}\rho^*_\mu
 ({\bf r}_c)$.
Thus in the corresponding $N$-dimensional space, the  partial width 
amplitude  of
a level  is simply the projection of  its corresponding eigenstate
 $\psi_\lambda$ on   the channel vector  $\phi_c$.  We remark
that the scalar product  in Eq. (\ref{PartWampProj}) 
(that will be used throughout this paper)
 is different from the original scalar product defined in the spatial  
region occupied by the dot.
 
 We define the channel correlation
 matrix $M$  of lead $i$ by
$M^i_{cc^\prime} = \overline{ \gamma_c^{i*} \gamma^i_{c^\prime}}$.
 Using the scalar product expression (\ref{PartWampProj}) 
for the partial width amplitude and the GE average
 $\overline{\psi^\lambda_\mu \psi^\lambda_{\mu^\prime}} = 
{1\over N} \delta_{\mu \mu^\prime}$,
we can rewrite $M$ in terms of the scalar product of the
 corresponding channels
\begin{eqnarray}\label{ChannelCM}
   M_{cc^\prime}^i=
   {1\over N}\langle\phi_c^i\mid\phi_{c^\prime}^i\rangle \;.
\end{eqnarray}
The norm of a channel 
vector is determined by the corresponding mean partial width through 
$\langle\phi_c^i\mid\phi_{c}^i\rangle=N\overline{\Gamma_c^i}$. Channels
$c,c^\prime$ are called uncorrelated if 
$\overline{\gamma_c^{i\ast}\gamma_{c^\prime}^i}=0$ and equivalent if
$\overline{\Gamma_c^i}=\overline{\Gamma_{c^\prime}^i}$. In general
 channels can
be correlated or non-equivalent and the channel vectors $\phi_c$ can 
therefore 
be non-orthogonal and have different norms. Channels in different leads are 
assumed to be uncorrelated. The matrices $M_{cc^\prime}^{l,r}$ thus
contain all the channel information relevant for the statistical description of 
the dot's conductance through Eqs. (\ref{PeakG}), (\ref{PartialW}) and 
(\ref{PartWampProj}).

\section{Conductance Correlations in Symmetric Dots}

 In experiments on quantum dots one can vary an external parameter $x$ such as
 the shape of the dot or the strength of an applied magnetic field and trace 
the corresponding change in the amplitude of a given conductance peak.
 We are then interested in calculating the conductance peak correlator 
(\ref{gParmCorr}), assuming the electron's dynamics inside the dot is 
chaotic (or disordered) at all values of $x$.  In this section we
show that the conductance correlator is universal upon an appropriate
 scaling of the parameter $x$ and we obtain its universal form. 

 \subsection{Gaussian Random-Matrix Process}

To calculate the correlation between conductance peaks
that belong to different parameter values it is necessary to use a
random-matrix model which incorporates a parametric dependence.
 Such a model,
the Gaussian random-matrix process (GP) which constitutes a natural 
generalization of Dyson's Gaussian ensemble (GE) \cite{DysonRMT} 
 has recently 
been proposed and applied to the calculation of universal 
parametric correlation functions in chaotic and disordered systems  \cite{GP}. 

 A GP is a 
set of $N\times N$ random matrices $H(x)$ whose elements satisfy
\begin{eqnarray}\label{GPmoments}
   \overline{H_{\lambda \sigma}(x)}&=&0 \;,\nonumber \\
   \overline{H_{\lambda \sigma}(x)H_{\mu \nu}(x^\prime)}&=&
   {a^2\over{2\beta}}f(x-x^\prime)g^{(\beta)}_{\lambda \sigma,\mu\nu} \;,
\end{eqnarray}
where 
   $g^{(\beta=1)}_{\lambda\sigma,\mu\nu}=\delta_{\lambda \mu}
\delta_{\sigma\nu}+\delta_{\lambda\nu}\delta_{\sigma\mu}$
and  $g^{(\beta=2)}_{\lambda\sigma,\mu\nu}= 2
\delta_{\lambda\nu}\delta_{\sigma\mu}$.
For situations with time-reversal invariance we have $\beta=1$ and $H(x)$ are
real symmetric matrices, whereas for broken time-reversal symmetry
 $\beta=2$ and $H(x)$ are complex Hermitian. The corresponding GPs are
 termed Gaussian orthogonal process (GOP) and Gaussian unitary
 process (GUP), respectively.  

 A GP is characterized by the short distance behavior of  its 
correlation function $f$
\begin{eqnarray}\label{fSD}
   f(x-x^\prime)\approx 1-\kappa\mid x-x^\prime\mid^\eta \;,
\end{eqnarray}
where $\kappa$ and $\eta$ ($0< \eta \le 2$) are constants. The
GP unfolded energies $\varepsilon_\lambda=E_\lambda/\Delta$
 (where $\Delta$ is the average level spacing) satisfy a diffusion law
 \cite{GP,GPlong} at short distances
\begin{eqnarray}\label{LevDiff}
   \overline{(\Delta\varepsilon_\lambda)^2}=D\mid\Delta x\mid^\eta
   +{\cal O}\left(\mid\Delta x\mid^{2\eta}\right)
\end{eqnarray}
where 
\begin{eqnarray}\label{DiffConst}
   D=\lim\limits_{\Delta x\rightarrow 0}
   {\overline{(\Delta\varepsilon_\lambda)^2}\over\mid\Delta x\mid^\eta}
   ={4N\kappa\over \pi^2\beta}
\end{eqnarray}
plays the role of the diffusion constant. 
Upon the parameter scaling
\begin{eqnarray}\label{ParmScal}
   x\rightarrow\bar{x}=D^{1/\eta}x
\end{eqnarray}
the correlation function $f(x-x^\prime)$ (\ref{fSD}) becomes
$f\approx 1-{\pi^2\beta\over 4N}\mid\bar{x}-\bar{x}^\prime\mid^\eta$, and
all correlators become
universal functions of $\mid\bar{x}-\bar{x}^\prime\mid^\eta$.
 This universality can also be demonstrated through Dyson's 
Brownian motion model \cite{GPFP}.
The $\eta=2$ GPs are the only class of processes that are continuously
 differentiable in the parameter and therefore the ones
  suitable  to describe most physical systems \cite{GPlong}. For $\eta=2$ the 
diffusion scaling (\ref{ParmScal}) reduces to the level velocity scaling 
$D=C(0)\equiv\overline{\left(\partial\varepsilon_\lambda/\partial x\right)^2}$,
 first introduced in \cite{SpectCorr}.

\subsection{The Conductance Parametric Correlator}

When the dot is symmetric under reflection around the $y$-axis we have
$\Psi_\lambda(-x,y)=\pm\Psi_\lambda(x,y)$, thus 
$\Gamma_{c\lambda}^l=\Gamma_{c\lambda}^r\equiv\Gamma_{c\lambda}$ 
and the peak conductance (\ref{PeakG}) becomes
\begin{eqnarray}\label{PeakGsymm}
   G={e^2\over h}{\pi\over 4kT}\Gamma_\lambda
   \;.                         
\end{eqnarray}     
We take the dot Hamiltonian $H(x)$ to be a member of the appropriate GP with
energies $E_\lambda(x)$ and eigenfunctions $\psi_\lambda(x)$. The 
conductance 
correlator (\ref{gParmCorr}) reduces in this case to the level width correlator
\begin{eqnarray}\label{wParmCorr}
   c_G(x-x^\prime)=c_\Gamma(x-x^\prime)=
   \overline{{\tilde\Gamma}(x){\tilde\Gamma}(x^\prime)}
   \;,\;\;\;\;\;
   {\tilde\Gamma}(x)={\Gamma(x)-\overline{\Gamma}\over
   \sqrt{\overline{\Gamma^2}-\overline{\Gamma}^2}}
\end{eqnarray}
where 
$\Gamma(x)=\sum\limits_{c=1}^\Lambda \mid\langle\phi_c
\mid\psi (x)\rangle\mid^2$
(see (\ref{PartWampProj})).  Here and in the following we omit the eigenvector 
label
$\lambda$ and the lead label $i$. It is clear from the discussion  in Section 
III.A that this correlator becomes universal under the diffusion scaling 
(\ref{ParmScal}),
for any given set of channel vectors $\phi_c$. Furthermore, due to the
invariance of the GOP (GUP) under an $x$-independent orthogonal (unitary)
 transformation, this correlator can depend only on
the correlation matrix $M$ in  (\ref{ChannelCM}). However, we will prove below 
the stronger result that the width correlator is also independent of $M$.

 Since $M$ is Hermitian and positive-definite,
 it can be 
diagonalized by a unitary transformation $U$ under which the channel vectors 
$\phi_c$ transform into an orthogonal set of eigenchannels $\bar{\phi}_c$
($\phi_c =\sum_{c^\prime} \bar{\phi}_{c^\prime} U_{c^\prime c}$) and
\begin{eqnarray}\label{NewChannelCM}
   \bar{M}_{cc^\prime}={1\over N}\langle\bar{\phi}_c\mid\bar{\phi}_{c^\prime}
   \rangle=w_c^2\delta_{cc^\prime}
\end{eqnarray} 
where $w_c^2=\langle\bar{\phi}_c\mid\bar{\phi}_c\rangle/N$ are the (positive)
 eigenvalues of $M$. 
  Since the total width (in each lead) is invariant under 
a unitary transformation, we have 
$ \Gamma = \sum_c \mid \gamma_c\mid^2 = \sum_c \mid \bar{\gamma}_c\mid^2$
 (where $\bar{\gamma}_{c} = \langle\bar{\phi}_c\mid \psi_\lambda \rangle$ 
are the partial widths to decay to the eigenchannels) and the width correlator
 thus depends only on the eigenvalues
$w_c^2$ of the correlation matrix.  Defining the normalized eigenchannels
$\mid\hat{\phi}_c\rangle=\left(\sqrt{N}w_c\right)^{-1}\mid\bar{\phi}_c\rangle$
we can express the width in terms of the partial widths to decay into an 
orthonormal set of channels
\begin{eqnarray}\label{GammaWc}
   \Gamma(x)=N\sum\limits_{c=1}^\Lambda w_c^2\hat{\Gamma}_c(x)
   \;,\;\;\;\;\;
   \hat{\Gamma}_c(x) \equiv \mid\langle\hat{\phi}_c\mid\psi(x)\rangle\mid^2 \;,
\end{eqnarray}
and the numerator of the width correlator  (\ref{wParmCorr}) becomes
\begin{eqnarray}\label{wPChat}
 \overline{\Gamma(x)\Gamma(x^\prime)}-
   \overline{\Gamma}^2  =
 \sum\limits_c w_c^4\left(
   \overline{\hat{\Gamma}_c(x)\hat{\Gamma}_c(x^\prime)}-
   \overline{\hat{\Gamma}_c}^2\right)+
   \sum\limits_{c\neq c^\prime} w_c^2w_{c^\prime}^2\left(
   \overline{\hat{\Gamma}_c(x)\hat{\Gamma}_{c^\prime}(x^\prime)}-
   \overline{\hat{\Gamma}_c}\;\overline{\hat{\Gamma}_{c^\prime}}\right)
  \;.
\end{eqnarray}
We now observe that the averaged quantities in (\ref{wPChat}) do not depend
on the subscripts $c,c^\prime$ but only on the orthogonality relation
$\langle\hat{\phi}_c\mid\hat{\phi}_{c^\prime}\rangle=0$. Indeed, if we chose a 
different pair $\hat{\varphi}_c,\;\hat{\varphi}_{c^\prime}$ satisfying
$\langle\hat{\varphi}_c\mid\hat{\varphi}_{c^\prime}\rangle=0$ there would 
exist a unitary transformation $U$ such that 
$\mid\hat{\varphi}_c\rangle=U\mid\hat{\phi}_c\rangle$ and
$\mid\hat{\varphi}_{c^\prime}\rangle=U\mid\hat{\phi}_{c^\prime}\rangle$ which
could be used to rotate $H(x)$ at each $x$ into $UH(x)U^\dagger$ with 
eigenfunctions $U\mid\psi_\lambda(x)\rangle$, leaving 
$\hat{\Gamma}_c(x),\;\hat{\Gamma}_{c^\prime}(x^\prime)$ unchanged. Since the
GP probability measure $P\left[H(x)\right] D[H(x)]$ (see \cite{GPlong})
 is invariant under global rotation, the 
corresponding averages remain unchanged as well. We then rewrite
 (\ref{wPChat}) as 
\begin{eqnarray}\label{wPChatS}
   \overline{\Gamma(x)\Gamma(x^\prime)}-
   \overline{\Gamma}^2 &=& \left(\sum\limits_cw_c^4 \right)
  \left[ \overline{\hat{\Gamma}_1(x)\hat{\Gamma}_1(x^\prime)}-
   \overline{\hat{\Gamma}_1}^2\right]   +
\left( \sum\limits_{c\neq c^\prime}w_c^2w_{c^\prime}^2\right)
  \left[ \overline{\hat{\Gamma}_1(x)\hat{\Gamma}_2(x^\prime)}-
   \overline{\hat{\Gamma}_1}\;\overline{\hat{\Gamma}_2} \right] \;, 
\end{eqnarray}
where the subscripts $1,2$ refer to any orthonormal pair 
$\hat{\phi}_c,\hat{\phi}_{c^\prime}$ ($c\neq c^\prime$).
 Now  the cross-channel
correlator (second correlator on the r.h.s. of  (\ref{wPChatS}))  is related to 
 the autochannel correlator (first term there) by
 \begin{eqnarray}\label{twocorr}
 \overline{\hat{\Gamma}_1(x)\hat{\Gamma}_2(x^\prime)}-
   \overline{\hat{\Gamma}_1}\;\overline{\hat{\Gamma}_2} = - 
{1\over N-1 }\left[
\overline{\hat{\Gamma}_1(x)\hat{\Gamma}_1(x^\prime)}-
   \overline{\hat{\Gamma}_1}^2\right] \;.
\end{eqnarray}
To prove (\ref{twocorr}) we use a complete basis of $N$ orthonormal channels  
$\hat{\phi}_i\;\;; i=1,\ldots,N$. Multiplying the two normalization conditions
$\sum\limits_{i=1}^{N} \mid \hat{\gamma}_{i}(x)\mid^2 =1$ and 
$\sum\limits_{j=1}^{N} \mid \hat{\gamma}_{j}(x^\prime)\mid^2 =1$
and averaging, we get
\begin{eqnarray}\label{normal}
N \overline{ \mid \hat{\gamma}_{1}(x)\mid^2 
\mid\hat{\gamma}_{1}(x^\prime)\mid^2}  +
N(N-1) \overline{ \mid\hat{\gamma}_{1}(x)\mid^2 
\mid\hat{\gamma}_{2}(x^\prime)\mid^2} =1 \;.
\end{eqnarray}
 Relation (\ref{normal}) leads immediately to (\ref{twocorr}).
According to (\ref{twocorr}) the cross-channel correlator 
 is negligible in the limit $N\rightarrow\infty$ compared with the
  autochannel correlator. From (\ref{wPChatS}) we obtain
\begin{eqnarray}\label{exact}
 \overline{\Gamma(x)\Gamma(x^\prime)}-
   \overline{\Gamma}^2 = \left(\sum\limits_cw_c^4 -{1 \over N -1} 
\sum\limits_{c\neq c^\prime} w_c^2w_{c^\prime}^2 \right)
  \left[ \overline{\hat{\Gamma}_1(x)\hat{\Gamma}_1(x^\prime)}-
   \overline{\hat{\Gamma}_1}^2\right] \;,
\end{eqnarray}
and therefore 
\begin{eqnarray}\label{wPCremarkable}
   c_\Gamma(x-x^\prime)=
   \overline{{\tilde{\hat{\Gamma}}}_1(x){\tilde{\hat{\Gamma}}}_1(x^\prime)}
   \;.
\end{eqnarray}

This result (\ref{wPCremarkable}) is remarkable in that the conductance
correlator in a symmetric dot is not only universal but also independent of
the number of channels in the leads, their associated mean decay widths and the
correlations among them; it is given by the correlator of a dot with
single-channel leads. The universal form of this function can be computed using
any GP \cite{GP}. We choose the simple process \cite{AW}
\begin{eqnarray}\label{SimpleGP}
   H(x)=H_1\cos x+H_2\sin x
\end{eqnarray}
which is an $\eta=2$ GP with a correlation function 
$f(x-x^\prime)=\cos(x-x^\prime)$. $H_1,H_2$ are independent
matrices belonging to the appropriate GE. Fig. \ref{cGamAnd}  shows the results 
for both 
the GOP and GUP. We find  that $c_\Gamma(x-x^\prime)$ is approximated
 very well by
\begin{eqnarray}\label{wPCfit}
   c_\Gamma(\bar{x}-\bar{x}^\prime)=\left[{1\over 1+(\bar{x}-\bar{x}^\prime)^2
   /\alpha_\beta^2}\right]^\beta 
   \;,
\end{eqnarray}
namely a Lorentzian in the orthogonal case and a squared Lorentzian in the
unitary case. Using 300 simulations with $N=150$ and estimating the 
statistical error at each $x$, a least-squares fit to (\ref{wPCfit}) gives 
$\alpha_1=0.48\pm 0.04$ and $\alpha_2=0.64\pm 0.04$.  The numerical 
simulations of the correlators (diamonds) and 
their best fit  to (\ref{wPCfit}) (dashed) are shown in Fig. \ref{cGamAnd}.
We have also confirmed in our simulations the independence
of the details of the channels by computing the correlator (\ref{wParmCorr})
with various sets of channel vectors $\phi_c$ and verifying the agreement with
(\ref{wPCfit}). Comparing the orthogonal and unitary results shows that 
breaking time-reversal symmetry accelerates the decorrelation. In
Section IV.C we use perturbation theory to derive an exact analytic
expression for $c_\Gamma$ to leading order in $\Delta \bar{x}$. This
leading order 
differs from that of (\ref{wPCfit}), and thus expression
(\ref{wPCfit}) is just an approximation. 

To test the universality of the conductance correlator, we model 
a disordered dot by the two-dimensional Anderson Hamiltonian 
\cite{AndersonModel,DisorderPT}
\begin{eqnarray}\label{Anderson}
   H=\sum\limits_\alpha(u_\alpha+w_\alpha)a_\alpha^\dagger a_\alpha
   -\sum\limits_{\langle\alpha\beta\rangle}e^{i\theta_{\alpha\beta}}
   a_\alpha^\dagger a_\beta
\end{eqnarray}
where  the second sum is
over nearest-neighbors lattice sites only.   The on-site energies $w_\alpha$  
are
distributed uniformly over the interval $[-W/2,W/2]$. The $u_\alpha$ describes 
 the non-random component of the potential whereas the phases 
$\theta_{\alpha\beta}$ are determined by the magnetic flux. 
We use a $n_x\times n_y=27\times 27$ lattice folded into a cylinder 
whose symmetry axis lies along $x$, imposing appropriate boundary conditions. 
The left and right leads are represented by arrays $A^{l,r}$ of $m_1\times m_2$
lattice points with total widths given by 
$\Gamma_\lambda^i=\sum\limits_{\vec{r}_\alpha\in A^i}(v_\alpha^i)^2
\mid\Psi_\lambda(\vec{r}_\alpha)\mid^2$ such that each point contact 
$\vec{r}_\alpha$ constitutes a channel with coupling $(v_\alpha^i)^2$ 
\cite{DotDistSUSY}. 

We introduce a parametric dependence by applying a step potential along $x$,
$u(x,y)=u_0\Theta(x)$, using its strength $u_0$ as the parameter. 
 The results for the conductance correlator, which are
presented in Fig.  \ref{cGamAnd} (left) for $2\times 2$  leads and 
 $v_\alpha^2=1$,  are in excellent agreement 
with the GOP prediction  (\ref{wPCfit}) (with $\beta=1$).

We break time-reversal symmetry by applying a magnetic field $B$ pointing along
the cylinder axis, corresponding to 
$\theta_{\alpha\beta}={2\pi\over n_y}\Phi/\Phi_0$ in (\ref{Anderson}) if
$\vec{r}_\alpha,\vec{r}_\beta$ are neighbors along $y$ and 
$\theta_{\alpha\beta}=0$ otherwise, where $\Phi/\Phi_0$ is the magnetic flux 
associated with $B$ in units of the flux quantum $\Phi_0=hc/e$. We take 
$\Phi/\Phi_0=1/4$ to achieve maximal symmetry breaking. The parametric 
dependence is introduced through a step potential and 
the results are presented in Fig.  \ref{cGamAnd} 
(right) using the same lead arrangements as above. The agreement with the GUP
prediction (\ref{wPCfit}) (with $\beta=2$)  is again very good. 

We remark that the partial width is analogous to 
the eigenfunction intensity at a fixed spatial point $\vec{r}$.
It follows that the parametric eigenfunction  intensity correlator  in a 
chaotic system
is identical to the  width correlator $c_\Gamma$ (\ref{wPCremarkable}).
Denoting by $\Psi^{x}(\vec{r})$ the eigenfunction at  $\vec{r}$ for a value 
$x$ of an external parameter, we have 
\begin{eqnarray}
\overline{\tilde{I}(x) \tilde{I}(x^\prime)} = c_\Gamma(x-x^\prime) \;,\;\;\; 
I(x) =\mid \Psi^x(\vec{r})\mid^2 \;.
\end{eqnarray}
 The eigenfunction intensities are measurable in microwave cavity experiments
\cite{Microwave1,Microwave2}
and our predictions for $c_\Gamma(x-x^\prime)$ can be tested there as well.
 
\subsection{Relation to the Parametric Overlap Correlator}

It is interesting to note that the result (\ref{wPCfit}) is identical (within
statistical errors) 
to our result for the universal form of the averaged parametric overlap 
\cite{GP} 
\begin{eqnarray}\label{ParmOver1}
   o(x-x^\prime)=\overline{\mid\langle\psi_\lambda(x)\mid\psi_\lambda(x^\prime)
\rangle \mid^2} \;.
\end{eqnarray}

 We now derive an exact relation between the width and overlap
 correlators (for finite $N$) and show that they are identical in the limit 
$N \rightarrow \infty$.
Choosing a complete set of vectors  $\hat{\phi}_i\;;\;\; i=1,\ldots,N$
 the overlap correlator can be written as
\begin{eqnarray}\label{Overcorr}
o(x-x^\prime)= \sum\limits_{i}\overline{\mid \hat{\gamma}_i(x)\mid^2 \mid 
\hat{\gamma}_i(x^\prime)\mid^2} +
\sum\limits_{i\neq j} \overline{\hat{\gamma}_i^\ast(x)\hat{\gamma}_j(x)
 \hat{\gamma}_i(x^\prime)\hat{\gamma}_j^\ast(x^\prime)}
\end{eqnarray}
 Since the various correlators in (\ref{Overcorr}) are independent of  the 
particular channel $i$ or the pair $i,j$ of orthogonal channels we can rewrite 
it as 
\begin{eqnarray}\label{twoCorr}
o(x-x^\prime)=N\left(\overline{ \hat{\Gamma}_1(x) \hat{\Gamma}_1(x^\prime)}
 -\overline{\hat{\Gamma}_1}^2\right)
+{N(N-1) \over 2} \left[ \overline{\hat{\gamma}_1^*(x)\hat{\gamma}_2(x) 
\hat{\gamma}_1(x^\prime)\hat{\gamma}_2^*(x^\prime)} + c.c.\right]
+{1\over N} \;.
\end{eqnarray}
To proceed we need to use a relation between the two parametric correlators
 on the r.h.s. of (\ref{twoCorr}),
the width correlator and the ``exchange'' correlator:
\begin{eqnarray}\label{exchange}
{1\over 2}  \left[ \overline{\hat{\gamma}_1^*(x)\hat{\gamma}_2(x) 
\hat{\gamma}_1(x^\prime)\hat{\gamma}_2^*(x^\prime)} + c.c.\right] =
\frac{\beta}{2}\frac{N}{N-1}
\left[\overline{ \hat{\Gamma}_1(x) \hat{\Gamma}_1(x^\prime)}
 -\overline{\hat{\Gamma}_1}^2\right] \;. 
\end{eqnarray}
 At $x=x^\prime$, this relation is easily derived using the results of Appendix
A, and for small $x-x^\prime \neq 0$, it can be proven perturbatively
 (see Appendix B).
At the end of this section  we provide a proof of this non-trivial relation
for any $x,x^\prime$. Using (\ref{exchange}) in (\ref{twoCorr}) we find
\begin{eqnarray}\label{oneCorr}
o(x-x^\prime)=N  { \beta N+2  \over 2} \left[\overline{ \hat{\Gamma}_1(x) 
\hat{\Gamma}_1(x^\prime)} -\overline{\hat{\Gamma}_1}^2\right]  +{1  \over  N}\;.
\end{eqnarray}
 Finally, using the GE relation $\overline{\hat{\Gamma}_1^2}- 
\overline{\hat{\Gamma}_1}^2 = 2(N-1) / N^2(\beta N +2)$
(see Appendix A),
we find the exact relation between the overlap and width correlator
\begin{eqnarray}\label{overCon}
o(x-x^\prime) = \left(1 - {1  \over  N} \right) c_\Gamma(x-x^\prime) + {1  \over
N}\;.
\end{eqnarray}
In the limit $N\rightarrow \infty$ these two correlators become identical.

 We now return to prove (\ref{exchange}). We first rederive the independence
of the width correlator of the matrix $M$ in a different way.
The channel vectors $\phi_c$ can be transformed by a linear
 transformation $\phi_c =\sum_{c^\prime} \hat{\phi}_{c^\prime} F_{c^\prime c}$
  to orthonormal channels  $\langle \hat{\phi}_c \mid \hat{\phi}_{c^\prime}
 \rangle= \delta_{cc^\prime}$. In terms of $F$, the original channel matrix is 
$\langle\phi_c\mid \phi_{c^\prime}\rangle=(F^\dagger F)_{cc^\prime}$. A simple
 solution for $F$ is $F=(N W)^{1/2}U$ where $U$ is the unitary matrix that 
diagonalizes $M$ and $W$ is the diagonal matrix composed of the eigenvalues 
$w_c^2$ of $M$ (see previous section).
In this case $FF^\dagger= N W$ is diagonal. However, we can take 
advantage of the freedom
to make a unitary transformation $R$ on the orthonormal eigenchannels
of the previous section to obtain another orthonormal set 
$\hat{\phi}_c$. This leads to $F=R(NW)^{1/2}U$ 
which in general has a non-diagonal $FF^\dagger=N RWR^\dagger$. 
Calculating the width correlator directly from the expression 
 $ \Gamma = \hat{\gamma^\dagger}FF^\dagger \hat{\gamma} $ for the total
width, we now find

\begin{eqnarray}\label{GWCorr}
 \overline{\Gamma(x)\Gamma(x^\prime)}
 - \overline{\Gamma}^2& = & \left\{\sum_c \left[(FF^\dagger)_{cc}\right]^2
\right\} \left(\overline{\hat{\Gamma}_1(x) \hat{\Gamma}_1(x^\prime)}
 - \overline{\hat{\Gamma}_1}^2\right) 
\nonumber \\ 
&  & + \frac{1}{\beta}\left\{\sum_{c\neq c^\prime}\mid 
 (FF^\dagger)_{cc^\prime}\mid^2  \right\}
  \left[ \overline{\hat{\gamma}_1^*(x)\hat{\gamma}_2(x) \hat{\gamma}_1
(x^\prime)\hat{\gamma}_2^\ast(x^\prime)} + c.c .
\right]  \nonumber \\
& &+\left\{\sum_{c\neq c^\prime} (FF^\dagger)_{cc}
(FF^\dagger)_{c^\prime c^\prime}\right\}
\left(\overline{\hat{\Gamma}_1(x) \hat{\Gamma}_2(x^\prime)} - 
\overline{\hat{\Gamma}_1}\overline{\hat{\Gamma}_2}\right) \;,
\end{eqnarray}
 The third term on the r.h.s. of (\ref{GWCorr})) is just the cross-channel 
correlator
and according to  (\ref{twocorr}) can be combined with the  autochannel
 width correlator (the first term on the r.h.s. of  (\ref{GWCorr})). 
 Notice that for the 
choice made of $F$ in the previous section
($FF^\dagger=W$ diagonal) the second term on the r.h.s. of
(\ref{GWCorr}) vanishes, unlike the present case. 
However, since one must recover the same width correlator
in (\ref{wPChatS}) and in (\ref{GWCorr}), we conclude that the first
 two correlators on the r.h.s. of
(\ref{GWCorr}) must be related. Indeed, if we assume relation (\ref{exchange})
 to hold, the two terms in (\ref{GWCorr}) combine to give   
\begin{eqnarray}\label{redGW}
 \overline{\Gamma(x)\Gamma(x^\prime)}- \overline{\Gamma}^2 =
\left[{\rm Tr} (FF^\dagger)^2 -{1 \over N-1} \left( ({\rm Tr} FF^\dagger)^2 - 
{\rm Tr} (FF^\dagger)^2\right) \right] \left[ \overline{\hat{\Gamma}_1(x) 
\hat{\Gamma}_1(x^\prime)} - \overline{\hat{\Gamma}_1}^2 \right] \;.
\end{eqnarray}
Since ${\rm Tr} (FF^\dagger)^2={\rm Tr} M^2 = \sum_c w_c^4$ and
${\rm Tr} (FF^\dagger)={\rm Tr} M$ , Eq. (\ref{redGW})
is then identical to (\ref{exact}).

\section{The Rotation Scaling}

Although all parametric correlators become universal upon the diffusion scaling
(\ref{ParmScal}), performing this scaling is not always feasible. In particular,
the scaling factor $D$ is not directly accessible in quantum dot experiments.
 As a result of the
Coulomb blockade suppression of tunneling, the energy levels probed as the gate
voltage is varied are the ground states for successive electron numbers rather 
than the excited states for a fixed electron number. Therefore one can measure 
the charging energy $e^2/C$ but not the level spacing $\Delta$ which is
necessary to determine $D$. While in principle it is possible to  measure
 the variation of the ground state energy of the
dot as a function of $x$ by tracing the peak amplitude, the mean level spacing 
 $\Delta$  can only be estimated theoretically from e.g. a Fermi gas model. 
Since the interaction and the dot's shape affect $\Delta$, it is difficult to
 extract accurate values for $D$ from the data. 

   In this section we introduce a new scaling
which is calculated from the eigenfunctions alone, or alternatively
from the level widths. This makes it possible to extract the scaled parameter 
directly from the conductance data  (see also Sec. VI.C),
 so that our predictions for
  $c_G(\bar{x}-\bar{x}^\prime)$  could be tested experimentally.

\subsection{Eigenfunction Rotation Rate and Parameter Scaling}

We consider the quantity 
$\overline{\mid\Delta\gamma_\lambda \mid^2} \equiv 
\overline{\mid\gamma_\lambda(x^\prime) - \gamma_\lambda(x) \mid^2}$ where 
$\gamma_\lambda(x)=\langle\phi\mid\psi_\lambda(x)\rangle$ is the partial 
level width and $\phi$ is a channel vector which we take for convenience to be
normalized. This quantity is independent of our choice for  $\phi$ since the 
GOP (GUP) is invariant under an $x$-independent  orthogonal (unitary) 
transformation  that will transform $\phi$ into another normalized vector. 
 From first-order perturbation theory we have
\begin{eqnarray}\label{DelGamma}
   \Delta\gamma_\lambda=\gamma_\lambda(x^\prime)-\gamma_\lambda(x)=
   \sum\limits_{\mu\neq\lambda}{\langle\psi_\mu(x)\mid H(x^\prime)\mid
   \psi_\lambda(x)\rangle\over E_\lambda(x)-E_\mu(x)}\gamma_\mu(x)
   +i\omega_\lambda(x)\gamma_\lambda(x)\Delta x \;,
\end{eqnarray}
where $\omega_\lambda(x)$ is an arbitrary real function that corresponds
 to the freedom
we have in choosing the phase of $\psi_\lambda$. We first make the usual choice
 $\omega_\lambda=0$, but we shall see below that another choice
 would be useful for computational purposes. 
The variance of $\Delta\gamma_\lambda$ is calculated in two steps.
First, we average over $H(x^\prime)$ keeping $H(x)$ fixed using the conditional
two-matrix distribution \cite{GP}
\begin{eqnarray}\label{2MCondDist1}
   P\left[H(x^\prime)\mid H(x)\right]=  
   P\left[H(x),H(x^\prime)\right] \;/\; P\left[H(x)\right] \nonumber \\
   \propto \exp\left\{-{\beta\over{2a^2}}{1\over 1-f^2}
   {\rm Tr}\left[H(x^\prime)-fH(x)\right]^2\right\}
\end{eqnarray}
with $f\equiv f(x-x^\prime)$.
  We can write the conditional averages as
\begin{eqnarray}\label{Condav}
\overline{ H^\prime_{\mu \lambda}- fH_{\mu\lambda}} &= &0 \nonumber \\
 \overline{ (H^\prime_{\mu \lambda} - fH_{\mu\lambda})
(H^\prime_{\sigma \eta}  -fH_{\sigma \eta})}
& = &{a^2 \over {2\beta}}  g^{(\beta)}_{\mu \lambda,\sigma\eta}  (1-f^2) \;. 
\end{eqnarray}
 Here and in the following we use the notation 
$H_{\mu\lambda}=\langle\psi_\mu\mid H\mid\psi_\lambda\rangle$
and $H^\prime_{\mu\lambda}=\langle\psi_\mu\mid H^\prime
 \mid\psi_\lambda\rangle$,
where matrix elements are calculated in the basis of the eigenstates of $H(x)$.
Primed quantities  take their value at  $x^\prime$ whereas 
non-primed quantities take their value at $x$. 
 To do the conditional averaging we 
write $\mid\Delta\gamma_\lambda\mid^2$ in the form
\begin{eqnarray}\label{DelGamma2}
  \mid \Delta\gamma_\lambda\mid^2=
   \sum\limits_{\mu\nu\neq\lambda}{
  \left(H_{\mu\lambda}^\prime -fH_{\mu\lambda}\right)
  \left(H_{\nu\lambda}^{\prime \ast}-fH_{\nu\lambda}^\ast \right)
   \over (E_\lambda-E_\mu)(E_\lambda-E_\nu)}
   \gamma_\mu\gamma_\nu^\ast 
\end{eqnarray}
 which is convenient since $P\left[H^\prime\mid H\right]$ depends on
$H^\prime-fH$ (note that $H_{\mu\lambda}=0)$. 
After averaging over $H^\prime$ we get
\begin{eqnarray}\label{DelGamma2Av1}
   \overline{\mid\Delta\gamma_\lambda\mid^2}=
   {a^2\over 2\beta}(1-f^2)\overline{\sum\limits_{\mu\neq\lambda}
   {\mid\gamma_\mu \mid^2\over(E_\lambda-E_\mu)^2}}
   \;.
\end{eqnarray}

Next we perform the average over $H$. Since the joint eigenvalue
 and eigenfunction
distribution factorizes in the GE \cite{RMTreviews} and 
$\overline{\mid\gamma_\mu\mid^2}$ is $\mu$-independent we have, using the 
short-distance expansion of $f(x-x^\prime)$ (\ref{fSD}), 
\begin{eqnarray}\label{DelGamma2Av2}
   \overline{\mid\Delta\gamma_\lambda\mid^2}=
   \beta \overline{\mid\gamma_\lambda\mid^2}NZ\kappa\mid\Delta x\mid^\eta
\end{eqnarray}
where
\begin{eqnarray}\label{Zdef}
   Z \equiv {a^2\over \beta N}\overline{\sum\limits_{\mu\neq\lambda}
   {1\over (E_\lambda-E_\mu)^2}}
   \;.
\end{eqnarray}
Dyson \cite{DysonBMM} obtained an exact result for a related quantity,
\begin{eqnarray}\label{ZDyson}
   Z_D={a^2\over \beta N}
   {1\over N}\overline{\sum\limits_{\lambda,\mu\neq\lambda}
   {1\over (E_\lambda-E_\mu)^2}}
   ={1-{1\over N}\over 2(\beta-1)} \;.
\end{eqnarray}
However, $E_\lambda$ in (\ref{Zdef}) is near the middle of the spectrum where
the level density is roughly constant whereas in (\ref{ZDyson}) it ranges
over the whole semi-circle, hence $Z\neq Z_D$.
It is more suitable for our purpose to unfold the levels and approximate the 
sum in (\ref{Zdef}) using the two-point cluster function $Y_2(\epsilon)$ 
\cite{RMTreviews},
\begin{eqnarray}\label{Z2ptCluster}
   Z={a^2\over \beta N}{1\over\Delta^2}\overline{\sum\limits_{\mu\neq\lambda}
   {1\over (\varepsilon_\lambda-\epsilon_\mu)^2}}
   \approx{4\over\pi^2\beta}\int\limits_0^\infty{d\varepsilon\over\epsilon^2}
   \left[1-Y_2(\varepsilon)\right]
   \;.
\end{eqnarray}
The integral diverges for $\beta=1$, as does Dyson's result (\ref{ZDyson}).
For $\beta=2$ it can be calculated using the simple Fourier representation
$Y_2(\varepsilon)=\sin^2(\pi\varepsilon)/(\pi\varepsilon)^2=
\int\limits_{-1}^1 dk e^{-2\pi ik\varepsilon}(1-\mid k\mid)$, giving $Z=2/3$ 
compared to Dyson's $Z_D=1/2$. 

Eq. (\ref{DelGamma2Av2}) can be rewritten (for $\Delta x\rightarrow 0$) as
 an amplitude diffusion law
\begin{eqnarray}\label{GammaDiff}
   \overline{\mid\Delta\gamma_\lambda\mid^2}\;/\;
\overline{\mid\gamma_\lambda\mid^2}
   =\beta R\mid\Delta x\mid^\eta         
   + \ldots
\end{eqnarray}              
in analogy with the level diffusion law (\ref{LevDiff}), where 
\begin{eqnarray}\label{GamDiffConst}
   R \equiv \lim\limits_{\Delta x\rightarrow 0}
   {1\over \beta \overline{\mid\gamma_\lambda\mid^2}}
   {\overline{\mid\Delta\gamma_\lambda\mid^2}\over\mid\Delta x\mid^\eta}
   \;.
\end{eqnarray}
In the GP we have $R=\kappa N Z$. Eq. (\ref{GammaDiff})
 motivates the parameter scaling
\begin{eqnarray}\label{RotScal}
   x\rightarrow\bar{x}_r=R^{1/\eta}x
\end{eqnarray}
which we call the rotation scaling as explained below. Under this scaling, just
like the diffusion scaling (\ref{ParmScal}), the process correlation function
(\ref{fSD}) becomes
\begin{eqnarray}\label{fCorr}
f \approx 1-{1\over NZ}\mid\bar{x}_r-\bar{x}_r^\prime\mid^\eta \;, 
\end{eqnarray}
and since all
correlators depend on $N(1-f)$ rather than on $N$ and $f$ separately \cite{GP},
we conclude that all correlators are universal functions of 
$\mid\bar{x}_r-\bar{x}_r^\prime\mid^\eta$. We remark that $R$ can also be
 written as $R = {\beta}^{-1}\lim\limits_{\Delta x\rightarrow 0}
   {\overline{\mid\Delta\tilde{\gamma}_\lambda\mid^2} / \mid\Delta x\mid^\eta}$
 where $\tilde{\gamma}_\lambda \equiv \gamma_\lambda /
 \sqrt{\overline{\mid\gamma_\lambda\mid^2}}$ is an ``unfolded'' amplitude
 in analogy with the unfolded level used in the definition (\ref{DiffConst}).  

 For the case of broken  time-reversal symmetry (GUP), the amplitude 
$\gamma_\lambda$ is generally complex when evaluated according to 
(\ref{DelGamma}) with $\omega_\lambda=0$. Note, however, that the 
eigenfunctions $\psi_\lambda(x)$
are determined up to an ($x$-dependent) phase, and we can choose this phase
such that the amplitude $\gamma_\lambda =\langle \phi \mid \psi_\lambda\rangle$
 is real for any $x$.   For this choice to
be consistent  we must fix $\omega_\lambda(x)$ such that 
$\omega_\lambda(x)\gamma_\lambda(x)\Delta x$ cancels the imaginary 
part of the sum 
on the r.h.s. of (\ref{DelGamma}), giving
\begin{eqnarray}\label{DelGammaPhase}
   \Delta\gamma_\lambda=
   \sum\limits_{\mu\neq\lambda}{{\rm Re}H^\prime_{\mu\lambda}\over
   E_\lambda-E_\mu}\gamma_\mu
\end{eqnarray}
which is real as well. Since only the real part of the matrix elements of $H$
 is used in (\ref{DelGammaPhase}), 
 the expression (\ref{GamDiffConst}) for $R$ is now replaced by 
\begin{eqnarray}\label{GamrealDiff}
   R \equiv \lim\limits_{\Delta x\rightarrow 0}
   {1\over\overline{\gamma_\lambda^2}}
   {\overline{(\Delta\gamma_\lambda)^2}\over\mid\Delta x\mid^\eta}
   \;.
\end{eqnarray}
In practice it is more convenient to use this definition, according to which
$\gamma_\lambda=\sqrt{\Gamma_\lambda}$ and is thus real.

 In most physical applications, the dependence of  $H$ on $x$ is analytic
and the GP must be differentiable, i.e  $\eta=2$ \cite{GPlong}. The
scaling factor is then given by
\begin{eqnarray}\label{Rphys} 
R= {1 \over \beta \overline{\mid\gamma_\lambda\mid^2}}
\overline{\mid{\partial\gamma_\lambda \over \partial x} \mid^2}\;,
\end{eqnarray}
which can be interpreted as
the eigenfunction rotation rate. To see this we note that as $x$ increases, the
eigenfunctions rotate in the Hilbert space and this rotation is characterized
by an anti-Hermitian operator $\Omega(x)$
\begin{eqnarray}\label{OmegaDef}
   {\partial\over\partial x}\mid\psi_\lambda\rangle=
   \Omega\mid\psi_\lambda\rangle
\end{eqnarray}
whose matrix elements in the basis $\psi_\lambda$ are given by
  $ \Omega_{\mu\lambda}=
   \langle\psi_\mu\mid\partial\psi_\lambda/\partial x\rangle=
(1-\delta_{\mu\lambda}) {\left(\partial H_{\mu\lambda}/\partial x\right)
   \over E_\lambda-E_\mu}$
(see (\ref{DelGamma})). The basis-independent quantity 
$N^{-1}\overline{{\rm Tr}(\Omega^\dagger\Omega)}$ is obtained
 from (\ref{OmegaDef})
with the help of a complete set of orthonormal vectors $\phi_\mu$
\begin{eqnarray}\label{TraceOmega2}
 {1 \over N} \overline{ {\rm Tr}(\Omega^\dagger\Omega)}=
   {1\over N}\sum\limits_{\lambda\mu}
   \overline{\mid\langle\phi_\mu\mid
   \partial\psi_\lambda/\partial x\rangle\mid^2}=R
\end{eqnarray}
exploiting the $\phi_\mu$-independence of the average in the sum in 
(\ref{TraceOmega2}) and the fact that 
$\overline{\gamma_\lambda^2}=1/N$. $R$ is therefore the variance of the
matrix elements of the scaled rotation rate matrix 
$\sqrt{N}\Omega(x)=\Omega(x)/\sqrt{\overline{\mid\gamma_\lambda\mid^2}}$, in
analogy with $D$ being the variance of the scaled level velocity
$(\partial E_\lambda/\partial x)/\Delta$. It also follows from (\ref{OmegaDef})
that $R$ is the norm of the eigenfunction derivative,
\begin{eqnarray}\label{DerivNorm}
   R=\overline{\langle\partial\psi_\lambda/\partial x\mid
   \partial\psi_\lambda/\partial x\rangle}
   \;.
\end{eqnarray}

  If the system 
obeys the GE statistics at each $x$, then
 $\overline{ \partial\gamma_\lambda/\partial x} = 0$ and
the quantity  $\overline{\mid \partial\gamma_\lambda/\partial x\mid^2}$
 in the definition of $R$
gives the variance  of $\partial\gamma_\lambda/\partial x$. 
 This should be contrasted with the level velocity scaling $D$,
 where the physical system can have a
 drift such that  $\overline{\partial \epsilon_\lambda/\partial x }\neq 0$
which has to be removed.
 
 While the values of the 
scaling factors $D$ and $R$ are system-dependent, the ratio $R/D$  is
universal and is given by
\begin{eqnarray}\label{UnivRatio}
   R/D= {\beta \over 4}\pi^2 Z 
   \;.
\end{eqnarray}
For the GUP $Z=2/3$  and 
\begin{eqnarray}\label{UnivRatioGUP}
   R/D = {\pi^2 \over 3} \;\;\;\;\;\;\; (GUP)\;.
\end{eqnarray}
If we use Dyson's $Z_D$ in (\ref{UnivRatioGUP}), we obtain $R/D=\pi/2$.
 The results from simulations  are in clear agreement with  $R/D=\pi^2/3$,
 as can be seen in Fig. 
\ref{ScalAnd}.

 It is important to note that  $R$  diverges in the orthogonal case
 (in the limit $\Delta x\rightarrow 0$) because of the divergence
of $Z$ in (\ref{ZDyson},\ref{Z2ptCluster}) and the rotation scaling,
as defined above, is valid for the unitary case only. However, 
 it is possible to fix this flaw as we 
discuss in the next section.

\subsection{Rotation Scaling in the GOP}

We now analyze the divergence of $R$ in the orthogonal case and show that it
is nevertheless possible to use it to perform the scaling.  We note that 
$\overline{\mid\Delta\gamma_\lambda\mid^2}/
\;\overline{\mid\gamma_\lambda\mid^2}$ is bounded from above by 4 and
 is therefore finite for any $\Delta x$. This is in contradiction to 
(\ref{GammaDiff}), 
according to which this quantity diverges for any fixed $\Delta x>0$. 
This indicates that while the perturbation
expansion (\ref{DelGamma}) is valid for a given member of the ensemble (for
$\Delta x$ sufficiently small), the ensemble average of the leading order term
diverges due to contributions from members with small level spacings. 
  Nevertheless,  since  
$\overline{\mid\Delta\gamma_\lambda\mid^2}/
\;\overline{\mid\gamma_\lambda\mid^2}$ is finite, we can define a scaling factor
$\tilde{R} (\Delta x)$ (in analogy with (\ref{GamDiffConst})) for any finite 
value of $\Delta x \neq 0$ by 
\begin{eqnarray}\label{Rtilde}
   \tilde{R} (\Delta x) \equiv {1 \over \beta \overline{\mid
\gamma_\lambda\mid^2}}
{\overline{\mid\Delta\gamma_\lambda\mid^2} \over\mid\Delta x \mid^\eta}\;.
 \end{eqnarray}              
 $\tilde{R}$ diverges (in the GOP) as $\Delta x\rightarrow 0$.  
From $R=\kappa NZ$, it is clear that this divergence  is caused by the small 
level spacings $s=(E_\lambda-E_\mu)/\Delta$ which are
distributed according to  $p(s)\propto s^\beta$ for small s \cite{Billiard}.  
Indeed $\int ds s^{\beta-2} $ converges near $s=0$ for $\beta > 1$ but diverges
 logarithmically for $\beta=1$.  The expression 
(\ref{Z2ptCluster}) for $Z$ can be regularized by introducing a small spacing
cut-off at $\varepsilon=\delta$. From  $Y_2(\varepsilon) 
\approx 1-\pi^2 \varepsilon/6 + {\cal O} (\varepsilon^3)$  \cite{RMTreviews}, 
we find
\begin{eqnarray}\label{Zdiverge}
   Z_\delta=-{2\over 3}\ln\delta+{\cal O}(1)
   \;.
\end{eqnarray}
 To regularize the perturbative expression  (\ref{DelGamma2Av2}) we examine
the conditional average for $\mid \Delta \gamma_\lambda\mid^2$ at fixed $H(x)$
(see (\ref{DelGamma2Av1}))
\begin{eqnarray}\label{DelGfx}
   \overline{\mid\Delta\gamma_\lambda\mid^2} \mid_{{\rm fixed}\;H(x)}=
   {1 \over 2} \sum\limits_{\mu\neq\lambda}
   {\mid\gamma_\mu\mid^2\over(\varepsilon_\lambda-\varepsilon_\mu)^2}
   \mid \Delta \bar{x}\mid^\eta\;,
\end{eqnarray}
written in terms of the unfolded energies and the scaled parameter. This 
expression is
well-defined for any given member $H(x)$ of the GE ensemble at $x$. However,
for a member with a level spacing that is too small, the approximation 
(\ref{DelGfx}),
 which is of the order $\mid\Delta x\mid^\eta$, breaks down unless we also make
$\mid\Delta x\mid^\eta$ smaller. Consequently, the average over all $H(x)$ at 
fixed $\mid\Delta x\mid^\eta$ leads to divergence.  To avoid that, we  use
a small spacing cut-off $\delta$ (as for $Z$ above) that depends on $\Delta x$.
 Noting that the parameter which determines the goodness of the
approximation (\ref{DelGfx}) is $\mid\Delta x\mid^\eta /\delta^2$, we make the
 conjecture that  a regularized perturbative expansion can be obtained
by assuming $\delta^2 \propto \mid\Delta\bar{x}\mid^\eta$. This is equivalent 
to relating two (unfolded) energy scale
  $\delta^2 \propto \overline{(\Delta \varepsilon)^2}$  where
the latter is associated with a change $\Delta \bar{x}$ in the scaled 
external parameter.  This regularization  leads to 
\begin{eqnarray}\label{RtildeDiv}
   {\tilde R}= \kappa N\left( - {1 \over 3}\ln\mid\Delta \bar{x}\mid^\eta
+{\cal O}(1) \right) \;.
\end{eqnarray}
 Our conjecture is supported by an exact result for 
$N=2$ which exhibits this behavior (see Section VII), and is confirmed 
numerically 
for large $N$ below. In analogy with the universal ratio $R/D$ for the GUP 
(\ref{UnivRatioGUP}) we now have a universal ratio 
for the GOP (for small $\Delta \bar{x}$)
 \begin{eqnarray}\label{UnivRatioGOP}   
  {{\tilde R} \over D}=-{\pi^2 \eta\over 12}\ln\left(\Delta \bar{x}\right)+
{\it Const.}  \;\;\;\;\;\;\; (GOP)\;.
\end{eqnarray}
 Eq. (\ref{UnivRatioGOP}) can be used to determine the
 scaling factor $D$ (in cases it is immeasurable) from the measured  
$\tilde{R}$.

Fig. \ref{ScalAnd} (left panel) presents the scaling ratio ${\tilde R}/D$
 computed with a
finite $\Delta x$ as a function of $\ln\Delta\bar{x}$. Shown are results of 
GP and Anderson model simulations compared to our conjecture for the orthogonal
case (\ref{UnivRatioGOP}). Similar calculations are shown (right panel) for 
the unitary
 case as a function of $\Delta x$ and they converge to (\ref{UnivRatioGUP}) in
the limit $\Delta x \rightarrow 0$. The agreement is very good.  We remark
that the smaller $\Delta x$ is, the more simulations should be performed to get
good statistics due to increasing fluctuations. Consequently, the GOP 
divergence would be difficult to observe experimentally.

\subsection{Perturbative Calculation of the Width Correlator}

We now calculate  $c_\Gamma(x-x^\prime)$ for a symmetric dot 
perturbatively to leading order.
A convenient way to calculate the width correlator is to use the relation
\begin{eqnarray}\label{ConDeltaG}
c_\Gamma(x-x^\prime)= 1- {1 \over 2} {\overline{(\Delta \Gamma_\lambda)^2} \over
\overline{ \Gamma_\lambda ^2} - \bar{\Gamma}_\lambda^2} \;,
\end{eqnarray}
where $\Delta \Gamma_\lambda \equiv \Gamma_\lambda(x^\prime)
 - \Gamma_\lambda(x)$. 
 Using  $\Gamma_\lambda(x^\prime)= \mid\gamma_\lambda(x^\prime)\mid^2$ in 
(\ref{DelGamma}) we get
\begin{eqnarray}\label{GammaPrime}
   \Gamma_\lambda^\prime=\Gamma_\lambda
   +\sum\limits_{\mu\neq\lambda}\left(
   {H^\prime_{\mu\lambda}\over E_\lambda-E_\mu}\gamma^\ast_\lambda\gamma_\mu
+c.c.\right)+ \ldots \;.
\end{eqnarray}
After averaging 
\begin{eqnarray}\label{GammaDel}
   (\Gamma_\lambda^\prime-\Gamma_\lambda)^2=
   \sum\limits_{\mu\nu\neq\lambda}
  { H^\prime_{\mu\lambda} H^{\prime \ast}_{\nu\lambda}
   \over(E_\lambda-E_\mu)(E_\lambda-E_\nu)}
\gamma_\mu\gamma^\ast_\nu\Gamma_\lambda + c.c.
\end{eqnarray}
over $H(x^\prime)$ while keeping $H(x)$ fixed, similarly to the calculation 
(\ref{DelGamma2}), we are left with
\begin{eqnarray}\label{GammaDel1}
   \overline{(\Gamma_\lambda^\prime-\Gamma_\lambda)^2}=
   {2 a^2\over\beta}(1-f^2)\overline{\sum\limits_{\mu\neq\lambda}
   {1\over(E_\lambda-E_\mu)^2}\Gamma_\mu\Gamma_\lambda}
\end{eqnarray}
where the average is over $H(x)$. Eqs. (\ref{fSD},\ref{Zdef},
\ref{GamDiffConst}) then give
\begin{eqnarray}\label{GammaDel2}
   \overline{(\Gamma_\lambda^\prime-\Gamma_\lambda)^2}=
   4\overline{\Gamma_\mu\Gamma_\lambda}R\mid\Delta x\mid^\eta
   \;.
\end{eqnarray}
Using the GE relation $\overline{\Gamma_\mu\Gamma_\lambda}={\beta\over 2}
 {N \over N-1}  \left(\overline{\Gamma_\lambda^2}-\overline{\Gamma_\lambda}^2
\right)$
(valid for $\mu\neq\lambda$) together with (\ref{ConDeltaG})  we  obtain in
the level velocity scaling
\begin{eqnarray}\label{wPCperturb}
   c_\Gamma(x-x^\prime)=1-\beta {N \over N-1} {R\over D}
\mid \bar{x}-\bar{x}^\prime\mid^\eta
   \;.
\end{eqnarray}

For the GUP we use Eq. (\ref{UnivRatioGUP}) to find
\begin{eqnarray}\label{wPCshort}
   c_\Gamma(x-x^\prime)\approx 1-{2\pi^2 \over 3}
   \mid \bar{x}-\bar{x}^\prime\mid^\eta \;
  \end{eqnarray}
as $N \rightarrow \infty$. When we compare (\ref{wPCshort}) with the
 leading-order behavior of the squared Lorentzian
 (\ref{wPCfit}) for the GUP case we find $\alpha_2=\sqrt{3}/\pi\approx 0.55$.
The discrepancy with the least-square fit value quoted below (\ref{wPCfit}) 
indicates that (\ref{wPCfit}), while being a good approximation, is not exact.
Indeed, including 
higher-order terms in the perturbative calculation introduces odd powers of
$\mid\bar{x}-\bar{x}^\prime\mid^{\eta/2}$ into $c_\Gamma$ (see Section
V.D). 

   In the GOP the perturbative approximation (\ref{GamDiffConst}) diverges.
Using a regularization similar to that invoked in the previous section amounts
to replacing $R$ in  (\ref{GamDiffConst}) by  $\tilde{R}$ as given  by 
(\ref{RtildeDiv}). 
 Using (\ref{UnivRatioGOP}) this leads to the following  short-distance 
behavior of the GOP conductance correlator 
\begin{eqnarray}\label{wPCperturbGOP}
   c_\Gamma(x-x^\prime) \approx 1 -  { N \over N-1}\mid\bar { x} -
\bar{x}^\prime\mid^\eta
\left( -{\pi^2\eta  \over 12} \ln \mid\bar{x}-\bar{x}^\prime\mid +Const. 
\right)  \;.
\end{eqnarray}
This non-analyticity near the origin indicates that the Lorentzian
(\ref{wPCfit}) which provides a very good approximation for 
$c_\Gamma(x-x^\prime)$
is not the exact result. We remark that the $N=2$ case exhibits the same
behavior (see Section VII).

\subsection{The  Rotation Scaling  as the RMS Width Velocity} 

 In Ref. \cite{GP} we have shown that the quantity
 $\overline{(\Delta \varepsilon_\lambda)^2}$,
which measures the diffusion of the unfolded energy levels in $x$, is a 
universal 
 function of  $(\Delta \bar{x})^{\eta}$  (even when the latter is not small).
 Similarly, one can consider the quantity $\overline{(\Delta 
\Gamma_\lambda)^2} / \overline{\Gamma}_\lambda^2$ 
where $\Delta \Gamma_\lambda \equiv \Gamma_\lambda(x^\prime) -
 \Gamma_\lambda(x)$ 
as a measure of the diffusion of the ``unfolded'' widths $\Gamma_\lambda
/\overline{\Gamma}_\lambda$.
 Based on our discussion from Section III.A, this must also be a universal 
function
 of $(\Delta \bar{x})^{\eta}$. We have computed this function for the GOP and 
GUP using the simple GP $\cos x H_1 + \sin x H_2$, and it is shown in Fig. 
\ref{DGamma2}.

   Using the perturbative expression (\ref{GammaDel2}), we find that for small 
$\Delta x$
\begin{eqnarray}\label{DelGD}
 {1 \over \bar{\Gamma}_\lambda^2} \overline{(\Delta \Gamma_\lambda)^2}=
   4{N \over N+\beta/2} \tilde{R} \mid\Delta x\mid^\eta  \approx
4 {\tilde{R} \over D}\mid\Delta \bar{x}\mid^\eta \;.
\end{eqnarray}
For the GUP we expect to find $\tilde{R}/ D \rightarrow
R/D= \pi^2/3$
in the limit $\Delta x \rightarrow 0$, while for the GOP $\tilde{R}/D$
 should have the logarithmic universal behavior (\ref{UnivRatioGOP})
 at small $\Delta x$. 

 Eq. (\ref{DelGD}) leads us to an alternative definition of the rotation scaling
\begin{eqnarray}\label{RotationG}
   \tilde{R}  \equiv {1 \over 4 \bar{\Gamma}_\lambda^2}
{\overline{\mid\Delta\Gamma_\lambda\mid^2} \over\mid\Delta x \mid^\eta}\;.
 \end{eqnarray}               
 In most physical applications, the parametric dependence is analytic
so $\eta=2$, and the scaling factor $\sqrt{R}$ is just the RMS of the rate
of change of the normalized width as a function of the parameter
\begin{eqnarray}\label{RotationGP}
 R = {1 \over 4 \bar{\Gamma}_\lambda^2}
\overline{\left( {\partial \Gamma_\lambda \over \partial x} \right)^2} \;.
\end{eqnarray}

   We remark that the definition (\ref{RotationGP}) of the rotation scaling 
has the advantage that there is no phase ambiguity in  $\Gamma(x)$.
Furthermore, in quantum dots experiments, the measurable quantity is the
width (i.e. conductance for a symmetric dot), so (\ref{RotationGP}) allows us
 to extract the scaled parameter directly from the data. 

 It is possible to calculate perturbatively higher order terms in 
 (\ref{DelGD}).
In the GUP, while the leading order converges, the next order does not.  For 
example, one of the terms of order $\Delta x ^{2 \eta}$ includes a diverging
 GUE average
$\overline{\sum\limits_{\mu \neq \lambda} 1/(E_\lambda - E_\mu)^4}$. If we 
introduce a small level spacing cutoff $\delta$, this average behaves as 
$1/\delta$. Using the same regularization as in Section IV.B, i.e. 
$\delta \propto \Delta \bar{x} ^{\eta/2}$, we find a perturbative term of order
$\Delta x ^{3\eta/2}$.
Carrying our regularization procedure to higher order terms we 
 make the following conjecture regarding the 
analytical behavior of (\ref{DelGD}) for finite $\Delta \bar{x}$. For the GOP
\begin{eqnarray}\label{GforGOP}
{\overline{(\Delta\Gamma_\lambda)^2} \over 4
\bar{\Gamma}_\lambda^2 }
=\mid\Delta\bar{x}\mid^\eta [h_1(\Delta \bar{x}) -h_2(\Delta\bar{x})
\ln \mid\Delta\bar{x}\mid^\eta] \;,
\end{eqnarray}
where $h_1$ and $h_2$ have power expansion in $(\Delta\bar{x})^\eta$.
For the GUP
\begin{eqnarray}\label{GforGUP}
{\overline{(\Delta\Gamma_\lambda)^2} \over 4
\bar{\Gamma}_\lambda^2 }
= \mid\Delta\bar{x}\mid^\eta [h_1(\Delta \bar{x}) -h_2(\Delta \bar{x})\mid 
\Delta \bar{x}\mid^{\eta/2}] \;,
\end{eqnarray}
where  $h_1$ and $h_2$ also have power expansion in $(\Delta \bar{x})^\eta$
(but are different from the corresponding functions for the GOP). 
Our conjecture is supported by the exact results for the $N=2$ GP derived
in Section VII. Note that  
for small $\Delta \bar{x}$ the functions in brackets in (\ref{GforGOP}) 
and (\ref{GforGUP}) are just $\tilde{R}/D$.
 
\subsection{Semiclassical Determination of the Rotation Scaling 
and the Correlation Field}

 To obtain a universal behavior it is necessary (for $\eta=2$)
 to scale the external parameter
by $\sqrt{D}$ or alternatively by $\sqrt{R}$. These scaling factors
are system-dependent and in principle could be determined
semiclassically. In this section we present a nonrigorous yet simple
semiclassical derivation of the rotation scaling factor $R$. The WKB
wavefunction inside the dot is given by $\Psi \sim A e^{i S/\hbar}$,
 where $S$ is the classical action.  In the limit $\hbar \rightarrow 0$,
the sensitivity of $\Psi$ to $x$ is due to the $x$-dependence of the action 
and we find, from (\ref{Rphys})
\begin{eqnarray}\label{Rotclass}
R \sim  {1 \over 2} \overline{\mid {1 \over \hbar} { \partial S \over \partial 
x}\mid^2} \;.
\end{eqnarray}

 In the important  case where $x$ is a magnetic field $B$,  it is possible to
calculate (\ref{Rotclass}) explicitly. In particular we would like
to derive an expression for the correlation field $B_c$ defined by
$\bar{B}_r = \sqrt{R} B \equiv B/B_c$.  The change in the action
when the field is changed by $\Delta B$ is given by
\begin{eqnarray}\label{DelS}
 \Delta S = {e \over c} \int \Delta {\bf A} \cdot d {\bf \ell} \;.
\end{eqnarray}
 
 In the following we shall distinguish between 
open and closed dots.  For {\it open} dots we write
$\Delta S =  {e \over c} {\Theta \over 2\pi} \Delta B$, where $\Theta$ 
is the area (times $2\pi$) swept by a classical trajectory.
 Using (\ref{Rotclass}) we then
find  
\begin{eqnarray}\label{RTheta}
R \sim  \overline{\Theta^2}/2\Phi_0^2 \;,
\end{eqnarray}
 where $\Phi_0= hc/e$ is the flux unit.  In a ballistic chaotic open dot 
the area distribution is given by 
 $P(\Theta) \propto e^{ - \alpha_{cl} \mid \Theta\mid}$
\cite{DotsSemiCl} 
so that $\overline{\Theta^2} = 2/\alpha_{cl}^2$ and 
$R = 1 /\alpha_{cl}^2 \phi_0^2 $. The correlation field is then given by
\begin{eqnarray}\label{Bcorr}
 {B}_c = \alpha_{cl} \Phi_0 \;.
\end{eqnarray}
 We note that  this is exactly the correlation field that  appears in the
 semiclassical derivation of the conductance correlator in ballistic
 open dots \cite{DotsSemiCl}. We also note 
that $\alpha_{cl} \propto \tau_{esc}^{-1/2}$ where $\tau_{esc}$ is the mean 
escape time from the dot \cite{Jensen91}. 

We now turn to the case of {\it closed} dots (the Coulomb blockade regime)
which is the one relevant to this paper.  For such dots the electron does not
escape but continues to accumulate phase until time $\tau=h/\Delta$, where 
$\Delta$ 
is the mean level spacing \cite{RobnikDots}. 
 Since $\Gamma \ll \Delta$, the decay time is much longer
than $\tau$, and it is that time scale $\tau$ that determines the correlation 
field.
 The action difference $\Delta S$ is then calculated for this time $\tau$. 
We write
\begin{eqnarray}\label{DStau}
\Delta S = {e \over c} \int_0^\tau dt \; \varphi (t) \;,
\end{eqnarray}
where $\varphi(t) \equiv \Delta {\bf A} (r(t)) \cdot {\bf v}(t)$.  To calculate 
$\overline{ (\Delta S)^2}$ we follow the method used in \cite{GOEtoGUE}
\begin{eqnarray}
\overline{ (\Delta S)^2} = 2{e^2 \over c^2} \int_0^\tau dt \; \int_0^{\tau-t}
 dt^\prime \;
\overline{ \varphi(t) \varphi(t+t^\prime)} \approx 2{e^2 \over c^2} 
\tau \int_0^\infty dt \;
\overline{ \varphi(0) \varphi(t)} \;. 
\end{eqnarray}
Since the magnitude of the velocity $v$ along the orbit is constant, the length 
$\ell$ travelled in time $t$ is given by $\ell= v t$. Measuring $\varphi$ as a 
function of $\ell$ and for an electron with unit velocity ($v=1$), we have
$\int_0^\infty dt\; \overline{ \varphi(0) \varphi(t)} = v \int_0^\infty d \ell 
\; \overline{ \hat{\varphi}(0) \hat{\varphi}(\ell)}$,
where $\hat{\varphi}(\ell) = \Delta {\bf A} \cdot \hat{\bf v} =
 {1 \over 2} \Delta B d(\ell)$ with $d$ being the distance of 
the corresponding orbit segment from the origin.  Defining a dimensionless
geometrical constant $\alpha_g$ \cite{GOEtoGUE}
\begin{eqnarray}
\alpha_g = {\cal A}^{-3/2} \int_0^{\infty} d\ell \; \overline{d(0)d (\ell)} \;,
\end{eqnarray}
we find from (\ref{Rotclass})
\begin{eqnarray}
R(\Delta B)^2 = \alpha_g \pi^2 \tau v {\cal A}^{-1/2}
 \left( {\Phi \over \Phi_0 }\right)^2 \;.
\end{eqnarray}
If we denote by $\tau_{cr} \equiv {\cal A}^{1/2} / v$ the time for the 
ballistic electron at the Fermi energy to cross the dot, then the 
correlation field  can be written as \cite{ExpCond}
\begin{eqnarray}\label{Bclosed}
B_c = { 1 \over  \pi \sqrt{\alpha_g}} \sqrt { {\tau_{cr} \over \tau}} 
{ \Phi_0 \over  {\cal A}}\;,
\end{eqnarray}
where 
\begin{eqnarray}\label{time}
{\tau \over \tau_{cr}} = { \sqrt{2mE{\cal A}} \over \hbar} \;,
\end{eqnarray}
and we have used $\Delta = 2\pi \hbar^2 / m {\cal A }$.  We note that
our  result
(\ref{Bclosed}) is similar to the one obtained for the field that is required
to  break time-reversal symmetry \cite{BR84,RobnikDots,GOEtoGUE}. 
From (\ref{Bclosed}) and (\ref{time}) we find that
 $B_c {\cal A} /\Phi_0 \propto {\cal N}^{-1/4}$ where
 ${\cal N}$ is the number of electrons in the dot.
 In the diffusive regime we expect the correlation field to be
given by a similar expression except that the ballistic crossing time is 
replaced by the time it takes the electron to diffuse across the dot 
$\tau_{cr} ={\pi \over 2} h / E_c$ where $E_c$ is the Thouless energy.
We then find
\begin{eqnarray}\label{Bcloseddif}
B_c \propto \sqrt { {\Delta \over E_c}} { \Phi_0 \over  {\cal A}} \;,
\end{eqnarray}
in agreement with the fluctuation correlation field obtained in the
 supersymmetry method \cite{SpectCorr},
and with the field strength that is required to break time reversal symmetry
 in disordered systems \cite{AIE93,EF94}.

  Finally we remark that for open dots the semiclassical conductance
correlator \cite{DotsSemiCl} can be written in terms of  the scaled field 
$\bar{B}_r$ as $c_g = \left(1 +\Delta \bar{B}_r^2\right)^{-2}$.  For open dots 
we find a similar form  
$c_g = \left(1 +\Delta \bar{B}_r^2/\alpha_{2r}^2 \right)^{-2}$
 (see Eq. (\ref{wPCfit}) for $\beta=2$)
but with $\alpha_{2r} \approx 1.15$ (where we have converted $\alpha_2$ of 
Section III.B  to its value $\alpha_{2r}$ in the rotation scaling).

\section{Universal Distributions of  the Width Velocity}

 In this section we restrict the discussion to the differentiable parametric
dependence ($\eta=2$) where
 the factors associated with the level velocity scaling and the rotation
 scaling are related to the second moment of the  level velocity and width 
velocity, respectively. One may be interested
in higher moments of these quantities or in general in their distribution.
 In fact, for any finite $\Delta x$ we can define an average  
velocity of the unfolded levels and widths with
 respect to the scaled parameter $\bar{x}$
\begin{eqnarray}\label{Rates}
 v \equiv {\Delta \varepsilon \over \Delta \bar{x}} \;; \;\;\;
r \equiv {1 \over \bar{\Gamma}} {\Delta \Gamma\over \Delta \bar{x}} \;. 
\end{eqnarray}
 Their respective distributions $P_{\Delta \bar{x}}(v)$ and
 $P_{\Delta \bar{x}}(r)$ are symmetric and universal for any given $\Delta 
\bar{x}$.
We now apply the perturbation techniques of Section IV to derive  expressions 
for these distributions in the limit of small $\Delta \bar{x}$.

 We first address the simpler case of the level velocity distribution
$P(v) = \overline{\delta \left(v -  \Delta \varepsilon /\Delta \bar{x}\right)}
=(2\pi)^{-1} \int_{-\infty}^\infty e^{ivt} 
\overline{e^{-i t \Delta \varepsilon / \Delta \bar{x}}}$.
  In first order perturbation theory  $v=(H^\prime_{\lambda\lambda} - 
H_{\lambda\lambda})/\Delta\bar{x}$,  and the conditional average
 on $H(x^\prime)$ at fixed $H(x)$ can be easily done by noticing that
$H_{\lambda\lambda}^\prime-f H_{\lambda\lambda}$ is a Gaussian 
variable with zero mean. Since cummulants higher than second vanish for a 
Gaussian variable, we find
\begin{eqnarray}\label{Condvel}
 \overline{e^{-i t \Delta \varepsilon / \Delta \bar{x}}}\mid_{{\rm fixed}\;H(x)}
= e^{-{1 \over 2} t^2 \overline{( \Delta \varepsilon / \Delta \bar{x})^2}}
=e^{-{1 \over 2} t^2} \;,
\end{eqnarray}
where we have used  $\overline{(\Delta \varepsilon / \Delta
 \bar{x})^2} = 1 +{\cal O} (\Delta \bar{x}^2)$ at fixed $H(x)$.
Since  the expression found in (\ref{Condvel}) is independent of $H(x)$,
 the average over the latter does not change it and
\begin{eqnarray}\label{Gaussv}
P(v) = (2\pi)^{-1/2} e^{-{1 \over 2} v^2} \;.
\end{eqnarray}
 Thus in the limit of small $\Delta \bar{x}$, the level velocity distribution
 is simply a Gaussian. In Fig. \ref{probvr} (top panel) we show (for both
 the orthogonal and
 unitary cases) the universal distribution (\ref{Gaussv}) (solid lines) and 
compare it
with $P_{\Delta \bar{x}}(v)$ for $\Delta \bar{x}=0.1$ (circles), as calculated 
from the GP.  Only at larger values $\Delta \bar{x}  > 0.2$ we observe 
deviations from the limiting distribution (\ref{Gaussv}). In Fig.
\ref{probvrd} (top panel) we show the universal form of the distributions
 $P_{\Delta \bar{x}}(v)$  for such larger values of $\Delta \bar{x}$.

We now turn to the more interesting calculation of  $P_{\Delta \bar{x}}(r)$ in 
the limit of small $\Delta \bar{x}$. To leading order in perturbation theory 
 (see (\ref{GammaPrime}))
$r= (\Delta\bar{x} \overline{\Gamma})^{-1}\sum\limits_{\mu\neq\lambda}
  \left( {H^\prime_{\mu\lambda}\over E_\lambda-E_\mu}
\gamma^\ast_\lambda\gamma_\mu +c.c. \right)$
 is a Gaussian variable at fixed $H(x)$ with zero 
conditional mean.  Therefore
\begin{eqnarray}\label{CondG1}
\overline{e^{-i t \bar{\Gamma}^{-1} \Delta \Gamma /
 \Delta \bar{x}}} \mid_{{\rm fixed} \; H(x)}
=  e^{-{1 \over 2} t^2 \bar{\Gamma}^{-2} \overline{ ( 
\Delta \Gamma /  \Delta \bar{x} )^2}} \;,
\end{eqnarray}
where the conditional average of the squared width's rate of change is
\begin{eqnarray}\label{CondG2}
 {1 \over  \bar{\Gamma}^2} \overline{\left( 
{\Delta \Gamma \over \Delta \bar{x}}\right)^2} \mid_{{\rm fixed}\; H(x)} =
2\sum\limits_{\mu\neq \lambda}{\Gamma_\mu \Gamma_\lambda / \bar{\Gamma}^2
\over (\varepsilon_\mu - \varepsilon_\lambda)^2} \;.
\end{eqnarray}
 Taking the Fourier transform of  (\ref{CondG1}), we can express $P(r)$ as
a GE average over Gaussian distributions
\begin{eqnarray}\label{DelGDist}
P(r) = {1\over 2} \overline{
\left[ \pi \sum\limits_{\mu\neq \lambda}{\Gamma_\mu
\Gamma_\lambda/\bar{\Gamma}^2 \over (\varepsilon_\mu - 
\varepsilon_\lambda)^2}\right]
^{-1/2} e^{-\left[4\sum\limits_{\mu\neq \lambda}{\Gamma_\mu
\Gamma_\lambda/\bar{\Gamma}^2 \over (\varepsilon_\mu - 
\varepsilon_\lambda)^2}\right]^{-1} r^2}}
\;.
\end{eqnarray}
 Thus we have reduced the process average in the original  expression
 for $P(r)$ to an ensemble average. Fig. \ref{probvr} (bottom panel)
 shows this limiting distribution (\ref{DelGDist})  by 
a solid line, while the  distribution $P_{\Delta\bar{x}}(r)$ for
 $\Delta\bar{x}=0.1$ is shown by circles. Notice that the variance of the 
limiting distribution $P(r)$ is just $R/D =\pi^2/3$ in the unitary case 
and $\tilde{R}/D$ in the orthogonal case.
In Fig. \ref{probvrd} (bottom panel) we show the universal distributions
 $P_{\Delta\bar{x}}(r)$  for larger values of $\Delta\bar{x}$. We use a 
logarithmic
 scale in order to show more clearly the deviations ar larger values of $r$ from
the limiting form (\ref{DelGDist}). 

We observe from Fig. \ref{probvrd} that for both the level and width velocity
 distributions the deviation from
 the corresponding limiting form occurs faster (with increasing 
$\Delta \bar{x}$) for the GUP than for the GOP. Also in each symmetry class
 (i.e. orthogonal or unitary)
the change in the distribution as $\Delta \bar{x}$ increases is slower for the
width velocity distribution than it is for the level velocity distribution. 

\section{ Conductance Correlations in Asymmetric Dots}

  In present experiments in the Coulomb blockade regime the dots do not have
reflection symmetry.
In this section we calculate the conductance correlator and its analytic 
short-distance behavior for such asymmetric dots.
We also express the rotation scaling in terms of RMS conductance velocity.

\subsection{Universal Correlators}

For dots without reflection symmetry $\Gamma_c^l\neq\Gamma_c^r$, and 
Eq. (\ref{PeakG}) should be used for the conductance $G$. If in addition the
leads are asymmetric 
we need in general two different channel correlation matrices 
$M_{cc^\prime}^l,\;M_{cc^\prime}^r$ of dimensions $\Lambda^l,\Lambda^r$ to 
characterize the dot, assuming that channels at different leads are
uncorrelated. However, Eq. (\ref{GammaWc}) still holds for $\Gamma^{l,r}$ 
separately and the conductance correlator $c_G(x-x^\prime)$ depends only on the 
eigenvalues $(w_c^2)^{l,r}$ of $M_{cc^\prime}^{l,r}$ as in the symmetric case,
since the invariance of the GP under global unitary transformations implies
that the choice of $\hat{\phi}_c^{l,r}$ does not affect averages involving
$\langle\hat{\phi}_c^{l,r}\mid\psi\rangle$.

 We first investigate $c_G(x-x^\prime)$ for an asymmetric dot with one-channel 
leads. This is the case which is relevant to present experiments in the Coulomb
blockade regime where one channel dominates in each lead \cite{ExpCond}.  
The strongest deviation from the symmetric dot  correlator is found for 
symmetric leads $\overline{\Gamma^l}=\overline{\Gamma^r}$.  The correlator for 
this case is calculated from simulations of the GP  (\ref{SimpleGP}) and
 presented in the top panel of Fig. \ref{cGasAnd} (dashed) and is compared 
  with the symmetric 
dot (\ref{wPCfit}) (dotted).  As in Section III.B we can also fit this 
correlator to (\ref{wPCfit})
but  now we find $\alpha_1= 0.37 \pm 0.04$ for the case of conserved
 time reversal symmetry (GOP)
and $\alpha_2= 0.54 \pm 0.04$ for broken time-reversal  symmetry (GUP).
 Also shown in  Fig. \ref{cGasAnd} are results from the Anderson model as 
described in Section III.B. In the top right panel  (broken time-reversal 
symmetry), the parameter dependence is introduced by
folding the cylinder into a torus and applying an additional magnetic field 
$B^\prime$ perpendicular to the torus plane. This amounts to setting 
$\theta_{\alpha\beta}={2\pi\over n_x}\Phi^\prime/\Phi_0\equiv\theta$ in 
(\ref{Anderson}) if $\vec{r}_\alpha,\vec{r}_\beta$ are neighbors along $x$ and 
using $\theta$ as the parameter.
 In the case of highly asymmetric leads, the small width
can be neglected in the denominator of Eq. (\ref{PeakG}), and $c_G$ reduces
again to the width correlator. We conclude that the conductance correlator
 for an asymmetric dot with single-channel leads varies between the two limiting
cases of symmetric leads and a symmetric dot.  We note that this variation is 
smaller in the unitary case.  We also find that  it takes a large
 asymmetry of the leads to see significant changes in the correlator.
 To study the effect of increased number of 
channels on the conductance correlator, we  computed $c_G(x-x^\prime)$
 for an asymmetric dot with multi-channels
 symmetric leads ($M^l=M^r$) assuming uncorrelated and equivalent channels.
  The GP results for
$8$ channels leads and the Anderson model results with $4\times4$ leads are
presented in Fig.  \ref{cGasAnd} (bottom) and show $c_G(x-x^\prime)$ approaching
the symmetric dot result. We emphasize that in this case the Anderson model
results are not expected to agree with the GP since their channel correlation
matrices differ, thus the agreement we find numerically suggests that the
correlator in the asymmetric case converges to a limiting form as the number
of channels increases and that this limit is the symmetric dot correlator. 

\subsection{Perturbative Expression for the Conductance Correlator}

 Using the techniques of Section IV.C,  we can calculate  $c_G(x-x^\prime)$ 
perturbatively (to leading order in $\Delta x$) for an asymmetric dot with 
arbitrary leads. 
We use the notation $G=(e^2 / h)(\pi / 2kT)g_\lambda$ where 
$g_\lambda \equiv\Gamma^l_\lambda \Gamma^r_\lambda/(\Gamma^l_\lambda+
\Gamma^r_\lambda)$ and 
define  $\Delta g_\lambda \equiv g_\lambda(x^\prime) - g_\lambda(x)$. 
For a small change in $x$ we have
\begin{eqnarray}\label{Delg}
\Delta g_\lambda = \left({\Gamma^l_\lambda \over \Gamma_\lambda}\right)^2
 \Delta \Gamma^r_\lambda + 
 \left({\Gamma^r_\lambda \over \Gamma_\lambda}\right)^2 
\Delta \Gamma^l_\lambda  \;,
\end{eqnarray}
where $\Gamma_\lambda \equiv \Gamma^l_\lambda + \Gamma^r_\lambda$ is
the total width of the level $\lambda$. Since the two leads are uncorrelated we
get
\begin{eqnarray}\label{Delg2}
\overline{(\Delta g_\lambda)^2 }= \overline{\left({\Gamma^l_\lambda \over 
\Gamma_\lambda}\right)^4
(\Delta \Gamma^r_\lambda)^2} + 
\overline{ \left({\Gamma^r_\lambda \over \Gamma_\lambda}\right)^4 
(\Delta \Gamma^l_\lambda)^2 } \;.
\end{eqnarray}

 Using  the same methods of Section III.B, and taking advantage of
 the invariance of  
$\Gamma^{l,r}_\lambda / \Gamma_\lambda$ under an $x$-independent
orthogonal (unitary) transformation, we find

\begin{eqnarray}\label{expDelg2}
 \overline{\left({\Gamma^{l,r}_\lambda \over \Gamma_\lambda}\right)^4 (\Delta 
\Gamma^{r,l}_\lambda)^2} = N^2 \left(\sum\limits_cw_c^4 -{1 \over N -1} 
\sum\limits_{c\neq c^\prime} w_c^2w_{c^\prime}^2 \right)
 \overline{\left({\Gamma^{l,r}_\lambda \over \Gamma_\lambda}\right)^4 (\Delta 
\hat{\Gamma}^{r,l}_{1\lambda})^2}  \nonumber \\
 \approx  {\rm Tr} (M^{l,r})^2
\overline{\left({\Gamma^{l,r}_\lambda \over \Gamma_\lambda}\right)^4 (\Delta 
\hat{\Gamma}^{r,l}_{1\lambda})^2}\;,
\end{eqnarray}
where $\hat{\Gamma}^{r,l}_{1\lambda}$ is the partial decay width to a single
normalized channel in the corresponding lead and the last equality
 is valid in the limit of large $N$.

 To calculate the expectation values in (\ref{expDelg2}) we
first perform the conditional average over $H(x^\prime)$  (keeping $H(x)$ 
fixed), for which we can use relation (\ref{GammaDel1}) to obtain 
\begin{eqnarray}\label{Delg2Pert}
\overline{(\Delta g_\lambda)^2 }= 4 \tilde{R} \left[  {\rm Tr} (M^{r})^2
\overline{\left({\Gamma^l_\lambda \over \Gamma_\lambda}\right)^4
 \hat{\Gamma}^r_{1\lambda} \hat{\Gamma}^r_{1\mu}} + 
 {\rm Tr} (M^{l})^2
\overline{\left({\Gamma^r_\lambda \over \Gamma_\lambda}\right)^4
 \hat{\Gamma}^l_{1\lambda} \hat{\Gamma}^l_{1\mu}} \right] \;,
\end{eqnarray}
where $\mu$ denotes a level different from $\lambda$.
Using the relation $c_G(x-x^\prime)= 1-  \overline{(\Delta g_\lambda)^2}
 / 2(\overline{ g_\lambda ^2} - \bar{g}_\lambda^2)$,  and  the independence of
the eigenfunctions $\lambda$ and $\mu$ (in the limit $N \rightarrow \infty$), we
obtain 
\begin{eqnarray}\label{Cperturb}
   c_G(x-x^\prime) \approx 1-  b_\beta 
{\tilde{R} \over D} \mid \bar{x}-\bar{x}^\prime\mid^2
   \;,
\end{eqnarray}
where the constant $b_\beta$ is given by
\begin{eqnarray}\label{Cconstant}
b_\beta = 2 { \overline{ g_\lambda^4 \left[  {\rm Tr} (M^{r})^2
 {\hat{\Gamma}_{1\lambda}^r \over (\Gamma_\lambda^r)^4} + 
{\rm Tr} (M^{l})^2 {\hat{\Gamma}_{1\lambda}^l \over (\Gamma_\lambda^l)^4} 
\right]} \over N (\overline{g_\lambda^2} - \bar{g}_\lambda^2)} \;.
\end{eqnarray}

 We remark that the quantities averaged over in (\ref{Cconstant}) are
functions of the independent (normalized) partial widths
 $\hat{\Gamma}_{c\lambda}^{l,r}$
through $\Gamma_\lambda^{l,r} =
 N \sum_c (w_c^{l,r})^2 \hat{\Gamma}_{c\lambda}^{l,r}$,
and that each partial width is distributed according to a $\chi^2$ distribution
in $\beta$ degrees of freedom with the same mean 
$\overline{\hat{\Gamma}_{c\lambda}^{l,r}}=1/N$. Therefore the constant
 $b_\beta$ is completely determined by the eigenvalues $(w_c^{l,r})^2$
and the number of channels in each lead.

 In the important case of a dot with one channel in each lead and symmetric
leads ($\bar{\Gamma}^l =\bar{\Gamma}^r$)  we find  $b_1=7/4$ and $b_2=3$.
It follows that for the GOP 
\begin{eqnarray}\label{CperturbGOP}
   c_G(x-x^\prime) \approx 1 - {7 \over 4} (\bar { x} -
\bar{x}^\prime)^2
\left( -{\pi^2  \over 6} \ln \mid\bar{x}-\bar{x}^\prime\mid +Const. 
\right)  \;,
\end{eqnarray}
 while for the GUP
 \begin{eqnarray}\label{wPCperturbGUP}
   c_G(x-x^\prime) \approx 1 -  \pi^2 (\bar { x} -\bar{x}^\prime)^2  \;.
\end{eqnarray}
  
\subsection{The  Rotation Scaling  as the RMS Conductance Velocity} 

 In asymmetric dots' experiments it is not possible to measure the width 
directly
and it is therefore useful to express the rotation scaling directly in terms of 
the conductance velocity rather than the width velocity (see Section IV.D). 
 This is accomplished through the 
 perturbative expression (\ref{Delg2Pert}) 
\begin{eqnarray}\label{DelCD}
 {1 \over \bar{G}_\lambda^2} \overline{(\Delta G_\lambda)^2} \approx
   2  \tilde{R}  b_\beta {\overline{g^2} - \bar{g}^2 \over \bar{g}^2} 
\mid\Delta x\mid^2 \;.
\end{eqnarray}

 Eq. (\ref{DelCD}) leads us to yet another definition of the rotation scaling
\begin{eqnarray}\label{RotationC}
   \tilde{R}  = r_\beta {1 \over \bar{G}_\lambda^2}
\overline{\left({\Delta G_\lambda \over \Delta x} \right)^2}\;,
 \end{eqnarray}  
where $r_\beta = \left[2 b_\beta \left( {\overline{g^2} - 
\bar{g}^2 \over \bar{g}^2} \right) \right]^{-1} $.

In the case of one-channel symmetric leads, we find
\begin{eqnarray}\label{RotationCGOP}
   \tilde{R}  \equiv {1 \over 7}{1 \over \bar{G}_\lambda^2}
\overline{\left({\Delta G_\lambda  \over \Delta x} \right)^2}
\end{eqnarray}  
 for the GOP, and
 \begin{eqnarray}\label{RotationCGUP}
 R = {5\over 24} {1 \over \bar{G}_\lambda^2}
\overline{\left( {\partial G_\lambda \over \partial x} \right)^2} 
\end{eqnarray}
for the GUP. 

    In practice, one can calculate  $R$  from the 
measured conductance peaks according to (\ref{RotationCGOP}) or
(\ref{RotationCGUP}), and then use the universal ratio $R/D$, as given by
Eq. (\ref{UnivRatioGOP}) or Eq. (\ref{UnivRatioGUP}),  to determine the 
diffusion scaling factor $D$ for the GOP or GUP,
 respectively.  The scaled parameter  $\bar{x}$ that leads to universal
 correlations is now easily determined from (\ref{ParmScal}).

\section{ The $N=2$ GP}

  For $N=2$ it is possible to calculate analytically the width correlator 
discussed
in this paper. While the results are different from the $N\rightarrow \infty$ 
limit, it is nevertheless instructive to study the $N=2$ case. In particular 
the analytical properties of  the rotation scaling factor and the
width correlator are similar.

\subsection{Two-matrix distribution}

 A GOE $2\times2$ matrix is real symmetric and thus has three independent real 
variables $H_{11}, H_{22}$ and $H_{12}$. We parametrize it in terms of its
 eigenvalues $E_1, E_2$
and the rotation angle $\alpha$ that characterizes the orthogonal $SO(2)$ 
matrix $R$ that
 diagonalizes $H$. Introducing the level spacing $E=E_1-E_2$  and average
 energy $\bar{E}=(E_1+E_2)/2$,  the GOE probability distribution is
${\cal P}_1(E,\bar{E},\alpha) \propto \mid E \mid 
e^{-\bar{E}^2/a^2} e^{-E^2/4a^2}$, 
where $-\infty < E, \bar{E} < \infty$ and $-\pi/2 < \alpha <\pi/2$.

 To solve for various two-point correlators, it is sufficient to calculate the 
joint two-matrix distribution $P(H,H^\prime) $ where $H \equiv H(x)$ and
$H^\prime \equiv H(x^\prime)$. 
Introducing the quantities $E, \bar{E}, \alpha$ and $E^\prime, 
\bar{E}^\prime,\alpha^\prime$ for $H$ and $H^\prime$, respectively, we find
\begin{eqnarray}\label{vi.2}
\lefteqn{ P_1(E, E^\prime, \bar{E}, \bar{E}^\prime,\alpha, \chi) = } \nonumber 
\\  &  & {\mid E E^\prime\mid \over 4(\pi a^2)^3 (1-f^2)^{3/2}}
e^{-(\bar{E}^2 + \bar{E}^{\prime 2} - 2f \bar{E}\bar{E}^\prime)/a^2(1-f^2)}
e^{-(E^2 + {E^\prime}^2 - 2f\cos 2\chi)/4a^2(1-f^2)}  \;,
\end{eqnarray}
 where $\chi=\alpha-\alpha^\prime$ is the rotation angle of  $R^{-1}R^\prime$.
Introducing polar coordinates  for $E,E^\prime$ through  
$E=\zeta \cos\theta, E^\prime=\zeta\sin\theta$ and integrating 
 over $\bar{E}, \bar{E}^\prime$ we obtain
\begin{eqnarray}\label{vi.3}
 P_1(\zeta, \theta, \alpha, \chi) = {1 \over 8 (\pi a^2)^2
 (1-f^2)} \zeta^3 \mid \sin2\theta\mid 
e^{-(1-f\sin2\theta \cos 2\chi)\zeta^2/4a^2(1-f^2)}\;,
\end{eqnarray}
where $0<\zeta < \infty, -\pi<\theta<\pi$ and $-\pi/2<\alpha, \chi < \pi/2$. 

 For wavefunction correlators that  are independent of the spectra, we can 
integrate over $\zeta$ and $\theta$.  
In order to keep the two levels at $x$ and $x^\prime$ 
in the same order  we require
 $E E^\prime > 0$, and therefore the integration range for $\theta$ is 
$0\leq \theta\leq\pi/2$ and $-\pi\leq \theta \leq -\pi/2$. We obtain
\begin{eqnarray}\label{vi.5}
&  & P_1( \alpha, \chi) = {2(1-f^2)\over \pi^2}  g_1(\xi) \;, \nonumber \\
&  & g_1(\xi)  =  {1 \over 1-\xi^2} \left( 1 
+ {2\xi\over \sqrt{1-\xi^2}} \arctan \sqrt{{1+\xi \over 1-\xi}} \right)
 \;;\;\;\; \xi \equiv   f\cos2\chi \;. 
\end{eqnarray}

  A $2\times 2$ GUE matrix  is characterized by two real variables
 $H_{11},H_{22}$ and one complex variable $H_{12}$. 
We parametrized it by its eigenvalues  $E_1,E_2$ and its diagonalizing
 unitary matrix 
\begin{eqnarray}\label{vi.6}
 U = \left( \begin{array}{cc}
 e^{i\phi} \cos \alpha e^{i\psi} &  e^{-i\phi}\sin\alpha e^{i\psi} \\
-e^{i\phi} \sin\alpha e^{-i\psi} &  e^{-i\phi}\cos\alpha e^{-i\psi}
\end{array} \right ) \;,
\end{eqnarray}
where $0\leq \alpha\leq \pi, 0\leq \phi \leq 2\pi, -\pi \leq \psi \leq \pi$.
We remark that the matrix $H$ is independent of $\psi$ but it is more convenient
 to use the parametrization (\ref{vi.6}) since it describes an element of the 
group $SU(2)$ whose invariant measure is $D[U]= \mid \sin 2\alpha\mid d\alpha
 d\phi d\psi$.  The GUE distribution is then given by
${\cal P}_2(E,\bar{E},\alpha,\psi,\phi) \propto E^2
 \sin 2\alpha e^{-2\bar{E}^2/a^2} e^{-E^2/2a^2}$.

To parametrize the two-matrix
distribution we introduce $E,\bar{E}, U$ and $E^\prime,\bar{E}^\prime, U^\prime$
 to describe $H$ and $H^\prime$, respectively. Instead of $U$ and $U^\prime$
 we then
use $U$ and ${\cal U} \equiv U^\dagger U^\prime$. Parametrizing ${\cal U}$ as in
(\ref{vi.6}) but with angles $\chi,\Phi,\Psi$ (replacing $\alpha, \phi,\psi$)
 and using the invariance of the group
 measure $D[U^\prime] = D[{\cal U}]$, we find
\begin{eqnarray}\label{vi.8}
\lefteqn{P_2 (E, E^\prime, \bar{E}, \bar{E}^\prime,\alpha, \chi) \propto} 
\nonumber \\ 
&  & E^2 {E^\prime}^2\mid  \sin 2\alpha \sin 2\chi \mid 
e^{-2(\bar{E}^2 + \bar{E}^{\prime 2} -2f\bar{E}\bar{E}^\prime)/a^2(1-f^2)} 
e^{-(E^2+{E^\prime}^2 -2fEE^\prime \cos 2\chi)/2a^2(1-f^2)}\;.
\end{eqnarray}
 Integrating over $\bar{E}, \bar{E}^\prime$, $\phi, \Phi$ and
$\psi,\Psi$, and using polar coordinates $\zeta, \theta$
in the $E,E^\prime$ plane, we find
\begin{eqnarray}\label{vi.9}
P_2(\zeta, \theta, \alpha, \chi) \propto
\zeta^5 \sin^2 2\theta\mid \sin 2\alpha \sin 2\chi \mid
e^{-\zeta^2(1-2f\sin 2\theta \cos 2\chi)/2a^2(1-f^2)} \;.
\end{eqnarray}
If $\zeta$ and $\theta$ are  integrated out  we have
\begin{eqnarray}\label{vi.11}
&  & P_2(\alpha, \chi)  \propto  \mid \sin 2\alpha \sin 2\chi \mid g_2(\xi) 
\nonumber \\
&   & g_2(\xi)  = {1 \over (1- \xi^2)^2} \left( {3 \over 2} \xi + {2\xi^2+1 
\over \sqrt{1-\xi^2}} \arctan\sqrt{{1+\xi \over 1-\xi}} \right) \;;\;\;\;
\xi  \equiv  f \cos 2\chi  \;.
\end{eqnarray}

\subsection{The Width Correlator}

 We take a channel $\phi=\left( \begin{array}{c} 1 \\ 0 \end{array} \right)$ and
the first eigenfunction $\psi(x)=\psi_1(x)$ so that 
for the GOP $\Gamma=R^2_{11}= \cos^2 \alpha$ and 
$\Gamma^\prime=R_{11}^{\prime 2}= \cos^2 \alpha^\prime
=\cos^2 (\alpha+\chi)$. The correlation between $\Gamma$
 and $\Gamma^\prime$ is then given by 
\begin{eqnarray}\label{vi.12}
 \overline{\Gamma \Gamma^\prime} = 
\overline{\cos^4 \alpha} \;\overline{\cos^2 \chi}  +
 {1 \over 4} \overline{\sin^2 2\alpha}\; \overline{\sin^2 \chi} -
 {1 \over 2} \overline{ \cos^2 \alpha \sin 2 \alpha}\; \overline{\sin{2 \chi} }
 =  {1 \over 8} (1+2\overline{\cos^2 \chi}) \;,
\end{eqnarray}
where we used $\overline{\cos^4 \alpha} = 3/8$ and
 $\overline{ \sin^2 2\alpha}=1/2$. Using $\overline{\Gamma^2}=3/8$, 
  we obtain for the width correlator 
\begin{eqnarray}\label{vi.13}
c_\Gamma= \overline{\cos 2\chi} \;.
\end{eqnarray}
 
 For the GUP  $\Gamma=\mid U_{11}\mid^2= \cos^2 \alpha$ and 
$\Gamma_1^\prime=\mid U_{11}^\prime\mid^2 = \cos^2 \alpha^\prime
= \mid \cos\alpha \cos \chi - \sin \alpha \sin \chi e^{-2i(\phi+\Psi)}\mid^2$,
and therefore
 \begin{eqnarray}\label{vi.14}
 \overline{\Gamma \Gamma^\prime}  &= &
\overline{\cos^4 \alpha} \; \overline{\cos^2 \chi}  +
 {1 \over 4} \overline{\sin^2 2\alpha} \; \overline{\sin^2 \chi} -
 {1 \over 2} \overline{ \cos^2 \alpha \sin 2 \alpha} \;\overline{ {\rm Re}
 e^{-2i(\phi+\Psi)}}\; \overline{\sin 2 \chi} \nonumber \\ 
&=&  (1+\overline{\cos^2 \chi})/6 \;,
\end{eqnarray}
where we have used $\overline{ \cos^4 \alpha} = 1/3$ and
 $\overline{ \sin^2 2\alpha}=2/3$. 
 Since $\overline{\Gamma^2}=1/3$ we obtain again Eq. (\ref{vi.13}) 
 for $c_\Gamma$.  

Eq. (\ref{vi.13}) can be calculated analytically for both classes. In the 
orthogonal case we have

\begin{eqnarray}\label{vi.15}
c_\Gamma^{\beta=1} =  {1-f^2 \over 4\pi} \int_0^{\pi/2}
 \cos2\chi [g_1(\xi) - g_1(-\xi)] d\chi = 
 {1-f^2 \over 4} \int_0^{\pi/2}{f \cos^2 2\chi \over (1-f^2\cos^2 2\chi)^{3/2}}
d\chi \;.
\end{eqnarray}
 The integral in (\ref{vi.15}) can be evaluated to give
\begin{eqnarray}\label{vi.16}
c_\Gamma^{\beta=1} =  {\pi \over 4} f(1-f^2) {}_2F_1(3/2,3/2;2,f^2)
 \;,
\end{eqnarray}
where  ${}_2F_1$ is a hypergeometric function.
 In the unitary case we find
\begin{eqnarray}\label{vi.17}
c_\Gamma^{\beta=2}  &=& {(1-f^2 )^{3/2}\over \pi} \int_0^{\pi/2}
 \sin 2\chi \cos2\chi [g_2(\xi) - g_2(-\xi)] d\chi  \nonumber \\  &=& 
 (1-f^2 ) \int_0^{\pi/2}{f \cos^2 2\chi \over (1-f^2\cos^2 2\chi)^{3/2}}
d\chi \;.
\end{eqnarray}
 Here too the integral can be done exactly to give
\begin{eqnarray}\label{vi.18}
c_\Gamma^{\beta=2} = 2 {\sqrt{1-f^2} \over \pi f} +2{2f^2-1 \over \pi f^2} 
\arcsin f \;.
\end{eqnarray}

\subsection{The width Correlator at Small Distances}

  We now examine the behavior of  
$\overline{(\Delta\Gamma)^2}/4\overline{\Gamma^2}=
(1-c_\Gamma)/2(\beta+1)$ for small $\Delta \bar{x} \equiv \bar{x}-
\bar{x}^\prime$
(see Eq. (\ref{DelGD}))
 by expanding the exact solution for $c_\Gamma$ in 
$1-f \approx  \beta\pi^2 (\Delta \bar{x})^2/8$.

For the GUP we find that (\ref{vi.18}) has a power expansion in $(1-f)^{1/2}$ 
and we
obtain the general form (\ref{GforGUP}) with $\eta=2$ suggested in Section 
IV.D where 
 \begin{eqnarray}\label{hforGUP}
&  & h_1(\Delta\bar{x}) = {\pi^2 \over 12} + {\pi^4 \over 32} (\Delta\bar{x})^2
 + \ldots \nonumber \\
&   & h_2(\Delta\bar{x}) ={ \sqrt{2} \pi^2 \over 9} + {29 \pi^4
 \over 360 \sqrt{2}}(\Delta\bar{x})^2 + \ldots \;.
\end{eqnarray}
 The leading order behavior agrees with the one derived  from
 perturbation theory.  We first use (\ref{UnivRatioGUP}) where for $N=2$ the 
relevant
$Z$ is Dyson's $Z_D=1/4$ (see Eq. (\ref{ZDyson})) to find $R/D=\pi^2/8$.
 From (\ref{DelGD}) we then obtain
\begin{eqnarray}
{\overline{(\Delta\Gamma)^2} \over 4\overline{\Gamma}^2 }
\approx  {\pi^2 \over 12} (\Delta \bar{x})^2 \;,
\end{eqnarray}
in agreement with the leading term in (\ref{hforGUP}).   

 For the GOP we find after expanding (\ref{vi.16}) the general from 
(\ref{GforGOP}) where 
\begin{eqnarray}\label{hforGOP}
&  & h_1(\Delta\bar{x}) = \pi^2{3\ln2 -1 -\ln(\pi^2/8) \over 64} +
 {\cal O}(\Delta\bar{x}^2) \nonumber \\
&   & h_2(\bar{x}) =  {\pi^2 \over 32}  + {3\pi^4 \over 1024} 
(\Delta\bar{x})^2 + {\cal O}(\Delta \bar{x}^4) \;.
\end{eqnarray}
 Thus the leading order  terms in $h_1$ and $h_2$ are constants, which is 
consistent with the large $N$ perturbative result (\ref{UnivRatioGOP})
 for $\tilde{R}/D$  (although with different constants).
 It is instructive to repeat the perturbative GOP calculation for the special 
case $N=2$
and see whether our regularization procedure for this case yields the correct 
expansion.
We obtain (\ref{DelGD}) with $\tilde{R}/D$ given by (\ref{UnivRatio}) but now
\begin{eqnarray}
Z = {2 \over \pi^2} \overline { 1 \over (\varepsilon_1 -\varepsilon_2)^2}
 = { 1\over 4} \int\limits_\delta^\infty
{1 \over \varepsilon} e^{-\pi^2 \varepsilon^2/16} d \varepsilon
 = -{1 \over 8}  {\rm Ei} \left( -{ \pi^2 \delta^2 \over 16}\right) \;,
\end{eqnarray}
where we introduce a small level spacing cutoff $\delta$ (in units of the mean 
spacing) and ${\rm Ei}$ is the exponential integral function. Using the identity
${\rm Ei} (-t) = {\cal C} + \ln t + \int\limits_0^t dt^\prime (e^{-t^\prime} 
- 1)/t^\prime$,
where ${\cal C}$ is the Euler constant, we find
\begin{eqnarray}
 Z =-{1 \over 8} \left[ {\cal C}  + \ln \left( \pi^2 \delta^2 \over 16\right) 
+ {\cal O} (\delta^2) \right] \;.
\end{eqnarray}
 Making the assumption $\delta^2 \propto (\Delta \bar{x})^2$, the quantity 
(\ref{DelGD}) is then given by 
\begin{eqnarray}
{\overline{(\Delta\Gamma)^2} \over 4\overline{\Gamma}^2 }
= (\Delta \bar{x})^2 \left[ -{\pi^2 \over 32} \ln (\Delta \bar{x})^2 + Const.
\right]  \;, 
\end{eqnarray}
in agreement with (\ref{GforGUP}) and the leading order in (\ref{hforGOP}).

\section{Conclusion}

In this paper we discussed the parametric correlation function
$c_G(x-x^\prime)$ of the conductance peaks in quantum dots in the regime of 
Coulomb blockade oscillations. Using the framework of the Gaussian process, we 
demonstrated its universality and obtained its functional form as well as its 
exact analytic behavior at short distances for
 both cases of conserved and broken time-reversal symmetry.   For a symmetric 
dot we  proved that $c_G$ is independent of the channel configuration, 
 while for asymmetric dots with multi-channel leads, our 
results indicate that the conductance correlator approaches its symmetric-dot 
form as the number of channels increases.
We proposed a new scaling procedure, the rotation scaling, which uses a scaling
factor that can be extracted directly from the conductance peaks and is 
therefore experimentally measurable. Our main result (\ref{wPCfit}) 
 can be readily tested experimentally.   It would also be interesting to 
apply our rotation scaling to chaotic systems other than quantum dots.

This work was supported in part by the Department of Energy Grant
 DE-FG02-91ER40608. We acknowledge C.M. Marcus for useful
 discussions.

\appendix
\section{Useful GE Relations}

 In this appendix we derive various useful partial width amplitudes correlations
in the Gaussian ensembles.  In the following $\phi_1,\phi_2$ denote
 two orthogonal and normalized channels, while $\psi_\lambda, \psi_\mu$
are two eigenfunction ($\lambda \neq \mu$).  $\gamma_{c\lambda} = \langle
\phi_c \mid \psi_\lambda\rangle$ are the partial width amplitudes.

 We first derive the distribution of  the partial width amplitudes  
$\gamma_{1 \lambda},\gamma_{2\lambda},\gamma_{1\mu},\gamma_{2\mu}$,
which is given by
\begin{eqnarray}\label{channelfuncdist}
P(\gamma_{1 \lambda},\gamma_{2\lambda},\gamma_{1\mu},\gamma_{2\mu})
= \int \prod\limits_{j=3}^N (d\psi_{j\lambda} d\psi_{j\mu})
\delta(\sum\limits_{i=1}^N \mid\psi_{i\lambda}\mid^2 -1)
 \delta(\sum\limits_{i=1}^N \mid\psi_{i\mu}\mid^2 -1)\delta(\sum\limits_{i=1}^N 
\psi_{i\lambda}^\ast\psi_{i\mu} )
\end{eqnarray}
 Using two sets of spherical coordinates for $\psi_{j\lambda}\;; \; j=3,\ldots,
N$ and $\psi_{j\mu}\;; \; j=3,\ldots,N$ in (\ref{channelfuncdist}) 
 we find 
\begin{eqnarray}\label{A.1}
P(\gamma) \propto \left[ (1-R_\lambda^2)(1-R_\mu^2) -
 \mid S \mid^2 \right]^{\beta {N-3 \over 2} -1} \;,
\end{eqnarray}
where $R_\lambda^2 = \mid\gamma_{1\lambda}\mid^2 
+\mid\gamma_{2\lambda}\mid^2$
and $S= \gamma_{1\lambda}^\ast\gamma_{1\mu} +
\gamma_{2\lambda}^\ast\gamma_{2\mu}$.
From  (\ref{A.1}) we can calculate various moments and correlations
 of the partial amplitudes.

  However, most  of the useful relations can be derived in a simpler way. 
From the exact distribution of  a single amplitude \cite{RMTreviews}, 
$P(\gamma_{1\lambda}) \propto 
( 1- \mid\gamma_{1\lambda}\mid^2)^{\beta{N-1 \over 2} -1}$,
we find

\begin{eqnarray}\label{A.2}
\overline{\mid\gamma_{1\lambda}\mid^4} = {1 \over N} {\beta +2 \over \beta N+2}
 \;.
\end{eqnarray}
  By squaring the relation $\sum\limits_{i=1}^{N} \mid\gamma_{i\lambda}\mid^2=
1$, taking an average and using  (\ref{A.2}) we find

\begin{eqnarray}\label{A.3}
\overline{\mid\gamma_{1\lambda}\mid^2\mid\gamma_{2\lambda}\mid^2}=
{1 \over N} {\beta \over \beta N+2}  \;.
\end{eqnarray}
Similarly, if we average the square the relation  $\sum\limits_{\lambda=1}^{N} 
\mid\gamma_{1\lambda}\mid^2=1$ and use (\ref{A.2}) we obtain
\begin{eqnarray}\label{A.4}
\overline{\mid\gamma_{1\lambda}\mid^2\mid\gamma_{1\mu}\mid^2}=
{1 \over N} {\beta \over \beta N+2} \;.
\end{eqnarray}
 To derive the correlation between the widths of two eigenfunctions to decay
into two orthogonal channels, we  use $\sum\limits_{i,j=1}^{N} 
\overline{\mid\gamma_{i\lambda}\mid^2 \mid\gamma_{j\mu}\mid^2}=1$ together
with (\ref{A.4})
\begin{eqnarray}\label{A.5}
\overline{\mid\gamma_{1\lambda}\mid^2\mid\gamma_{2\mu}\mid^2}=
{\beta(N-1) + 2 \over \beta N+2}{1 \over N(N-1)}  \;.
\end{eqnarray}
 
Finally, by averaging the square of the modulus of  the
 orthogonality relation of the two eigenvectors 
$\sum\limits_{i=1}^N \gamma^\ast_{i\lambda}\gamma_{i\mu}=0$, 
and using relation (\ref{A.4}), we find
\begin{eqnarray}\label{A.6}
\overline{\gamma^\ast_{1\lambda}\gamma_{1\mu}
 \gamma_{2\lambda}\gamma^\ast_{2\mu}}
= -{1 \over N-1} \overline{\mid\gamma_{1\lambda}\mid^2\mid\gamma_{1\mu}\mid^2}
= - {\beta \over \beta N +2} {1 \over N(N-1)}\;. 
\end{eqnarray}

\section{Perturbative Calculation of the Cross-channel and Exchange Correlators}
  
 To prove (\ref{twocorr}) to leading order in perturbation theory
 we consider the quantity
$\overline{\left(\Gamma_{1\lambda}^\prime-\Gamma_{1\lambda}\right)
           \left(\Gamma_{2\lambda}^\prime-\Gamma_{2\lambda}\right)}$
where the subscripts $1,2$ correspond to an orthogonal pair of normalized 
channel vectors $\phi_{1,2}$, and 
$\Gamma_{i\lambda}(x)=\mid \gamma_{i\lambda}(x) \mid^2=
 \mid \langle\phi_i\mid\psi_\lambda(x)\rangle \mid^2$, $i=1,2$. Using 
(\ref{DelGammaPhase}) we have
\begin{eqnarray}\label{GammaDel12}
 (\Gamma_{1\lambda}^\prime-\Gamma_{1\lambda})
   (\Gamma_{2\lambda}^\prime-\Gamma_{2\lambda}) = 
  \left(\sum\limits_{\mu\neq\lambda}
   {H^\prime_{\mu\lambda}\over E_\lambda-E_\mu}\gamma^\ast_\lambda\gamma_\mu
+c.c \right)
 \left(\sum\limits_{\nu\neq\lambda}
   {H^\prime_{\nu\lambda}\over E_\lambda-E_\nu}\gamma^\ast_\lambda\gamma_\nu
+c.c \right) \;,
\end{eqnarray}
which upon averaging gives 
\begin{eqnarray}\label{GammaDel12av}
   \overline{(\Gamma_{1\lambda}^\prime-\Gamma_{1\lambda})
   (\Gamma_{2\lambda}^\prime-\Gamma_{2\lambda})}=
   2\left({\overline{\gamma^\ast_{1\lambda}\gamma_{2\lambda}
\gamma_{1\mu}\gamma^\ast_{2\mu}} + c.c.}\right)
   R\mid\Delta x\mid^\eta
   \;.
\end{eqnarray}

 The cross-channel correlation is then calculated from
 (\ref{GammaDel12av}) using
\begin{eqnarray}\label{CCDelG}
\overline{\Gamma_{1\lambda}\Gamma_{2\lambda}^\prime} - 
\overline{\Gamma_{1\lambda}} \;
\overline{\Gamma_{2\lambda}^\prime} = \overline{\Gamma_{1\lambda}
\Gamma_{2\lambda}}
 - \overline{\Gamma_{1\lambda}}\; \overline{\Gamma_{2\lambda}}  -
  \overline{(\Gamma_{1\lambda}^\prime
-\Gamma_{1\lambda})
   (\Gamma_{2\lambda}^\prime-\Gamma_{2\lambda})}/2 \;.
\end{eqnarray}
 After also using the GE  relation  $\left(
\overline{\gamma^\ast_{1\lambda}\gamma_{2\lambda}
\gamma_{1\mu}\gamma^\ast_{2\mu}} + c.c\right)/ 2= {\beta \over 2} {N \over N-1}
\left(\overline{\Gamma_{1\lambda}\Gamma_{2\lambda}} - 
\overline{\Gamma_{1\lambda}} \;\overline{\Gamma_{2\lambda}} 
\right)$ we get
\begin{eqnarray}\label{CCPer}
\overline{\Gamma_{1\lambda}\Gamma_{2\lambda}^\prime} -
 \overline{\Gamma_{1\lambda}}\;
\overline{\Gamma^\prime_{2\lambda}} = \left(\overline{\Gamma_{1\lambda}
\Gamma_{2\lambda}}
 - \overline{\Gamma_{1\lambda}}\; \overline{\Gamma_{2\lambda}} \right)
\left( 1 - \beta {N \over N-1} R \mid \Delta x\mid^\eta\right) \;.
\end{eqnarray}

Comparing with the autochannel correlator (\ref{wPCperturb}) and using the
 GE relation  $\overline{\Gamma_1 \Gamma_2}-
   \overline{{\Gamma}_1}\;\overline{{\Gamma}_2} = -  {1\over N-1}
\left(\overline{{\Gamma}_1 {\Gamma}_1}-
   \overline{{\Gamma}_1}^2\right)$,  we recover (\ref{twocorr}) to order
 $\mid\Delta x\mid^\eta$.

Relation (\ref{exchange}) can be similarly proven
 in perturbation theory.
 Writing the correlator of interest in the form
$ \left(\overline{\gamma^{\ast}_{1\lambda} \gamma_{2\lambda}
\gamma^{\prime \ast}_{1\lambda} \gamma^\prime_{2\lambda}} + c.c. \right)/2=
\overline{ \Gamma_{1\lambda} \Gamma_{2\lambda}} - 
 \overline{\mid \gamma^\ast_{1\lambda}\gamma_{2\lambda} - 
 \gamma^{\prime \ast}_{1\lambda}\gamma^\prime_{2\lambda} \mid^2}/2$,
we calculate the  second term on the r.h.s. 
Here we have to distinguish between the GOP and the GUP.
For the GOP
\begin{eqnarray}\label{GOPer}
 \overline{\mid \gamma^\ast_{1\lambda}\gamma_{2\lambda} - 
 \gamma^{\prime \ast}_{1\lambda}\gamma^\prime_{2\lambda} \mid^2}=
 2 \left( \overline{ \gamma_{1\lambda}^2 \gamma_{2\mu}^2} +
\overline{ \gamma_{1\lambda}\gamma_{2\lambda}\gamma_{1\mu}\gamma_{2\mu}}
\right)
R \mid \Delta x\mid^\eta \;,
\end{eqnarray}
while for the GUP we find
\begin{eqnarray}\label{GUPer}
 \overline{\mid \gamma^\ast_{1\lambda}\gamma_{2\lambda} - 
 \gamma^{\prime \ast}_{1\lambda}\gamma^\prime_{2\lambda} \mid^2}=
 4 \overline{ \mid\gamma_{1\lambda}\mid^2 \mid \gamma_{2\mu}\mid^2}
R \mid \Delta x\mid^\eta \;.
\end{eqnarray}
  Using also the GOE relation
 $\overline{ \Gamma_{1\lambda} \Gamma_{2\mu}} +
\overline{ \gamma_{1\lambda}\gamma_{2\lambda}\gamma_{1\mu}\gamma_{2\mu}}
 ={N \over N-1} \overline{\Gamma_{1\lambda} \Gamma_{2\lambda}}$
and  the GUE relation $\overline{ \Gamma_{1\lambda} \Gamma_{2\mu}}=
{N \over N-1} \overline{\Gamma_{1\lambda} \Gamma_{2\lambda}}$, we obtain

\begin{eqnarray}\label{exPer}
{1 \over 2} \left(\overline{\gamma^\ast_{1\lambda} \gamma_{2\lambda}
\gamma^{\prime \ast}_{1\lambda} \gamma^\prime_{2\lambda}} + c.c. \right)
= \overline{\Gamma_{1\lambda}\Gamma_{2\lambda}}
\left( 1 - \beta {N \over N-1} R \mid \Delta x\mid^\eta\right) \;.
\end{eqnarray}
  Comparing (\ref{exPer}) with (\ref{wPCperturb}) and using 
 $\overline{\Gamma_{1\lambda}\Gamma_{2\lambda}} = {\beta \over 2}
{N \over N-1} \left( \overline{ \Gamma_{1\lambda}^2} -
\overline{\Gamma_{1\lambda}}^2\right)$,
we recover (\ref{exchange}) to leading order in $\mid\Delta x\mid^\eta$.

\begin{figure}[p]
\caption{
Universal form of the conductance correlator for a symmetric dot  
(\protect\ref{wParmCorr}).  Shown are GP simulation results (diamonds) obtained
from (\protect\ref{SimpleGP}) and their best fit to (\protect\ref{wPCfit}) 
(dashed).  The circles are results from the Anderson model in a 
cylindrical  geometry for  leads of $2 \times 2$  point contacts, and
with the parameter being the strength of an external potential. In the
case of broken time reversal symmetry (right) a magnetic flux of 
$\Phi/\Phi_0=1/4$ is applied. For comparison we also plot the GUP
 result on the left and the GOP on the right (dotted). 
 The GP results are obtained using $300$
 simulations with $N=150$, taking the middle third of
the states, while the Anderson model results are from the middle $200$
 eigenfunctions of the Hamiltonian (\protect\ref{Anderson}) with $W=1$ on
 a $27\times 27$ lattice.
}
\label{cGamAnd}
\end{figure}

\begin{figure}[p]
\caption{
GP (diamonds with statistical error) and Anderson model (circles)
simulation results for the scaling factors ratio ${\tilde R}/D$ computed with
finite $\Delta x$, compared to the 
prediction in the orthogonal (\protect\ref{UnivRatioGOP}) and unitary
(\protect\ref{UnivRatioGUP}) cases. Notice that the ratio is plotted as a
function of $\ln\Delta\bar{x}$ for the orthogonal case and as a function of
$\Delta \bar{x}$ for the unitary case.  Here $\bar{x}=\protect\sqrt{D}x$ 
is the diffusion scaling. The Anderson model results correspond to the 
cases shown in Fig. \protect\ref{cGamAnd} but for $4\times4$ leads.
}
\label{ScalAnd}
\end{figure}

\begin{figure}[p]
\caption{ The universal form of $\overline{(\Delta \Gamma)^2}/\bar{\Gamma}^2$
as a function of $(\Delta \bar{x})^2$ for the GOP (left) and the GUP (right). 
The 
results are obtained from simulations of the process (\protect\ref{SimpleGP}).
The asymptotic value for large $\Delta \bar{x}$ is $4/\beta$.
}
\label{DGamma2}
\end{figure}

\begin{figure}[p]
\caption{ Top: the  level velocity distribution $P(v=\partial \varepsilon/ 
\partial \bar{x})$  given by (\protect\ref{Gaussv}) (solid line). Shown by 
circles is 
the distribution $P(v=\Delta\varepsilon/ \Delta\bar{x})$ calculated for 
 $\Delta \bar{x}=0.1$.
Bottom: The  width velocity distribution $P(r= \bar{\Gamma}^{-1}\partial \Gamma
/\partial \bar{x})$ given by  (\protect\ref{DelGDist}) 
(calculated  for $\Delta \bar{x} = 0.01$). Also shown is the 
distribution $P(r= \bar{\Gamma}^{-1}\Delta \Gamma /\Delta \bar{x})$ for
 $\Delta \bar{x}=0.1$ (circles). 
}
\label{probvr}
\end{figure}

\begin{figure}[p]
\caption{ Top: the average level velocity distribution $P(v)$ (where $v=\Delta
\varepsilon/ \Delta\bar{x}$) for various values of $\Delta \bar{x}=$ 0.25
(squares), 0.5 (pluses), 0.75 (crosses) and 1 (diamonds). The solid line is
the limiting Gaussian distribution (\protect\ref{Gaussv}).
Bottom: the average width velocity distribution $P(r)$ 
(where $r= \bar{\Gamma}^{-1}
\Delta \Gamma/\Delta \bar{x} $) for the same values of $\Delta \bar{x}$ as for
 $P(v)$.
The solid histogram is the limiting distribution (\protect\ref{DelGDist}). 
Notice that we use a logarithmic scale for
$P(r)$ in order to distinguish between the various cases at larger values of 
$r$.
}
\label{probvrd}
\end{figure}

\begin{figure}[p]
\caption{
Universal form of the conductance correlator (\protect\ref{gParmCorr}) in 
asymmetric dots.
 Top: GP simulation results with single-channel symmetric leads (dashed) 
and Anderson model results with $1\times 1$ leads (pluses and crosses). 
Bottom: GP simulation 
results with $8$ equivalent channels in each lead (dashed) and Anderson model
results with $4\times 4$ point-contact leads (squares and circles). 
The Anderson model is as described in  Fig. \protect\ref{cGamAnd}  where
the parametric dependence is on the strength of an external potential, 
except  in one case (crosses) where  the parametric dependence 
is that of an additional external magnetic field in a toroidal geometry.
For comparison we also plot  the correlator for a symmetric dot
(\protect\ref{wPCfit})  (dotted).
}
\label{cGasAnd}
\end{figure}

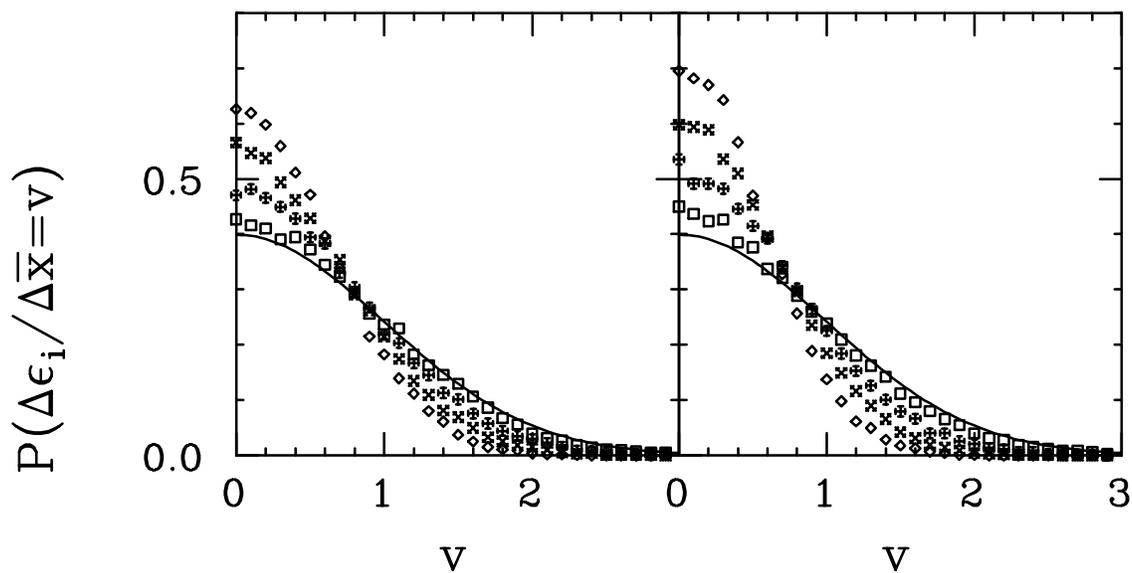
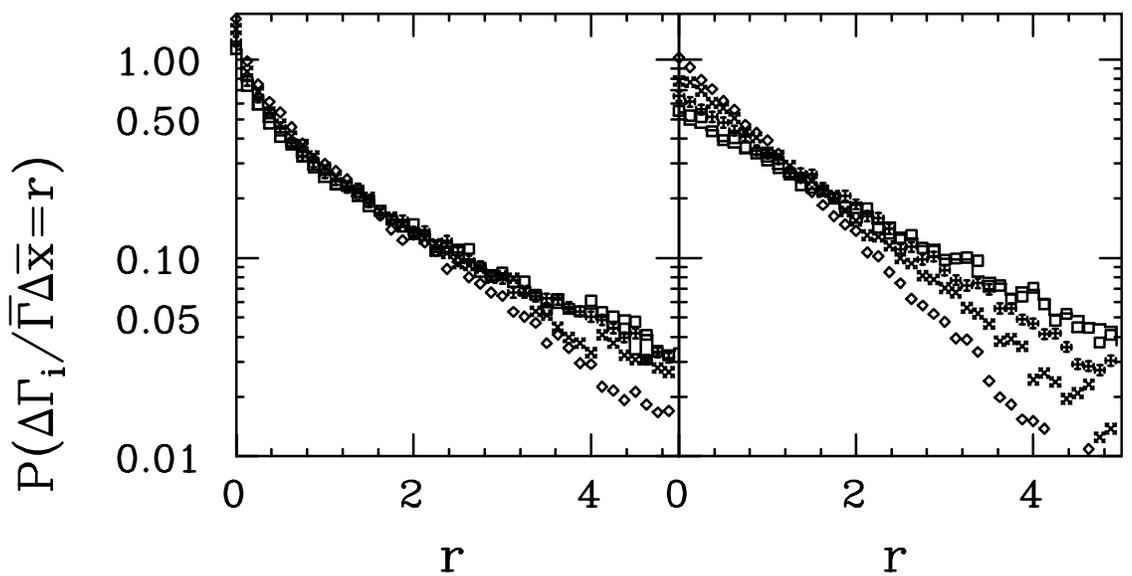
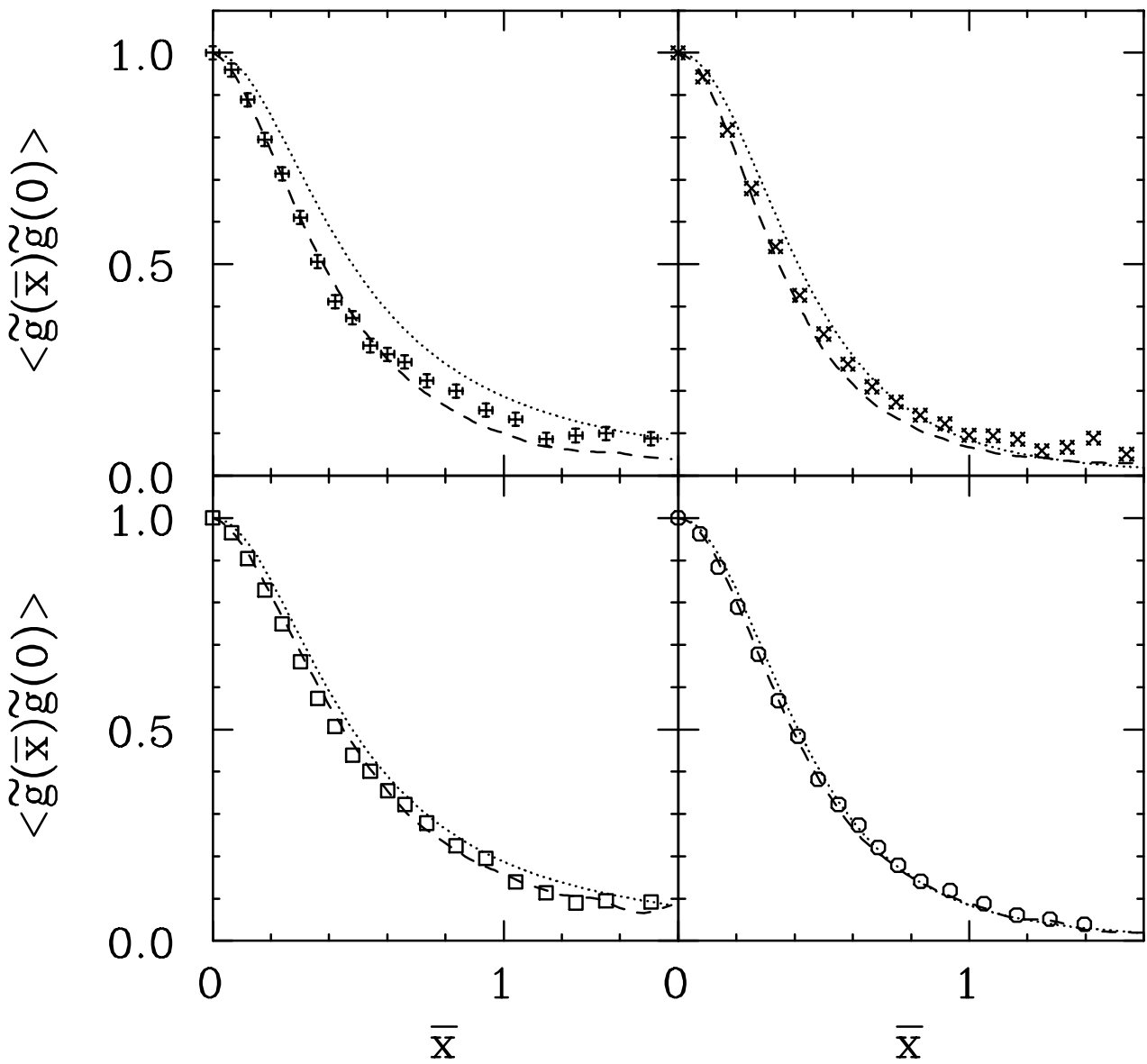
\end{document}